\documentclass{aa}  
\usepackage{graphicx}
\usepackage{rotating}
\usepackage{placeins}
\usepackage{txfonts}
\begin{document} 
   \title{Hot gas heating via magnetic arms in spiral galaxies
   \thanks{Based on observations obtained with {\sl XMM-Newton}, an ESA science mission with instruments and contributions directly funded by ESA Member States and NASA}
   \fnmsep\thanks{All reduced radio and X-ray images (FITS format) are available in electronic form at the CDS via anonymous ftp to cdsarc.u-strasbg.fr (130.79.128.5)
or via http://cdsweb.u-strasbg.fr/cgi-bin/qcat?J/A+A/}
    }
   \subtitle{The case of M\,83}

\author{M. We\.zgowiec\inst{1}
\and M. Ehle\inst{2}
\and M. Soida\inst{1}
\and R.-J. Dettmar\inst{3}
\and R. Beck\inst{4}
\and M. Urbanik\inst{1}}
\institute{
Obserwatorium Astronomiczne Uniwersytetu Jagiello\'nskiego, ul. Orla 171, 30-244 Krak\'ow, Poland, \\
\email{markmet@oa.uj.edu.pl}
\and European Space Agency, European Space Astronomy Centre (ESA/ESAC), Camino Bajo del Castillo s/n, 28692 Villanueva de la Ca\~nada, Madrid, Spain
\and Ruhr-University Bochum, Faculty of Physics and Astronomy, Astronomical Institute, 44780 Bochum, Germany
\and Max-Planck-Institut f\"ur Radioastronomie, Auf dem H\"ugel 69, 53121 Bonn, Germany}
\offprints{M. We\.zgowiec}
\date{Received; accepted date}

\titlerunning{Hot gas heating...}
\authorrunning{M. We\.zgowiec et al.}

 
  \abstract
   {Reconnection heating has been considered as a potential source of the heating of the interstellar medium. In some galaxies, significant polarised radio
   emission has been found between the spiral arms. This emission has a form of `magnetic arms' that resembles the spiral structure of the galaxy. 
   Reconnection effects could convert some of the energy of the turbulent magnetic field into the thermal energy of the surrounding medium, leaving more ordered 
   magnetic fields, as is observed in the magnetic arms.}
   {Sensitive radio and X-ray data for the grand-design spiral galaxy M\,83 are used for a detailed analysis of the possible interactions of magnetic fields with 
   hot gas, including a search for signatures of gas heating by magnetic reconnection effects.}
   {Magnetic field strengths and energies derived from the radio emission are compared with the parameters of the hot gas calculated from the model fits 
   to sensitive X-ray spectra of the hot gas emission.}
   {The available X-ray data allowed us to distinguish two thermal components in the halo of M\,83. We found slightly higher average temperatures 
   of the hot gas in the interarm regions, which results in higher energies per particle and is accompanied by a decrease in the energy density of the magnetic fields.}
   {The observed differences in the energy budget between the spiral arms and the interarm regions suggest that, similar to the case of another spiral galaxy NGC\,6946, we
   may be observing hints for gas heating by magnetic reconnection effects in the interarm regions. 
   These effects, which act more efficiently on the turbulent component of the magnetic field, are expected to be stronger in the spiral arms. However, with the present data it is only possible 
   to trace them in the interarm regions, where the star formation and the resulting turbulence is low.}

    \keywords{galaxies: individual: M\,83 -- 
              galaxies: ISM -- 
              galaxies: spiral -- 
              galaxies: magnetic fields --
	      magnetic reconnection --
	      X-rays: galaxies
              }

   \maketitle
%

\section{Introduction}
\label{intro}

As demonstrated in a review by \citet{beck04}, there is a strong indication that the
magnetic field controls most of the dynamical processes in the interstellar medium (ISM). 
This is done mainly by the action of the magnetic pressure.
Heating by reconnection was already condsidered as a potential source of ISM heating
\citep[e.g.][]{raymond92,tanuma99, reynolds99},
but a lack of sensitive X-ray data of high spectral resolution meant that this hypothesis
could not be verified as the observational signatures were not clear.
Contemporary reviews of physics of the interstellar medium list photo-ionisation as heat sources
as well as ionisation by cosmic rays, the photoelectric effect on grain surface, or H$_2$ formation
on grains or shock heating. At low number densities of the ISM (below 0.1\,cm$^{-3}$),
the magnetic reconnection could also be a significant source of heating of the ISM.
As the magnetic reconnection acts more efficient in the galactic spiral arms due to stronger field tangling, it may remain undetected because both the heating and field
(dis)ordering are dominated by turbulence. In the interarm regions, however,
where the magnetic field is highly ordered, these effects may dominate turbulence heating and become easier to observe.

In NGC\,6946, large-scale ordered magnetic fields were found exactly between the star-forming spiral arms \citep{beck07}. The origin of 
these magnetic arms remains a topic of research. The large angular extent of this galaxy allowed for a detailed analysis of the properties of the hot gas 
in the interarm regions, that is, at the positions of the magnetic arms, with the use of sensitive X-ray observations \citep{wezgowiec16}. 
A slight increase in temperature of the hot gas was found when compared to the spiral arms. Higher energies of the hot gas particles were 
accompanied by the lower energies and higher regularities of the magnetic fields. These findings suggest that the reconnection heating was observed 
in the interarm regions of NGC\,6946. 

In this paper, we present the results of a similar analysis of the radio and X-ray data performed for M\,83, which is another grand-design face-on spiral galaxy that also shows prominent 
magnetic arms in the sensitive radio maps. Our findings, together with the results obtained for NGC\,6946, could provide arguments 
that the heating by magnetic reconnection should be included in contemporary models of the interstellar medium.

M\,83 or NGC\,5236 is the nearest barred spiral galaxy. The resulting large angular extent combined with low inclination (Table~\ref{astrdat}) 
makes it one of the spiral galaxies that are best studied at all wavelengths. The CO observations revealed a very high gas mass density, 
especially in its nuclear part \citep{lundgren04}, that leads to an intense starburst. M\,83 contains also a significant amount of molecular 
hydrogen \citep{israel01} and the observations with the seven dish Karoo Array Telescope \citep[KAT-7;][]{carignan13} revealed an envelope of the neutral hydrogen, 
that extends far beyond the star forming disc \citep{heald16}. This envelope is a rich reservoir to fuel the galactic star formation. 

\begin{table}[ht]
        \caption{\label{astrdat}Basic astronomical properties of M\,83}
\centering
                \begin{tabular}{lc}
\hline\hline
Morphological type \tablefootmark{a}& SAB(s)c	\\
Inclination	   & 15\degr	\\
Diameter D$_{25}$  & 13\farcm5	\\
R.A.$_{2000}$	   & 13$^{\rm h}$37$^{\rm m}$01$^{\rm s}$	\\
Dec$_{2000}$	   & -29\degr 51\arcmin 56\arcsec	\\
Distance\tablefootmark{b} & 4.66\,Mpc	\\
Column density $N_{\rm H}$\tablefootmark{c}& 3.96$\times$10$^{20}$\,cm$^{-2}$\\
\hline
\end{tabular}
\tablefoot{
All data except type, distance, and column density taken from HYPERLEDA database -- http://leda.univ-lyon1.fr -- see \citet{makarov14}.\\
\tablefoottext{a}{Taken from \citet{israel80}.}
\tablefoottext{b}{Taken from \citet{tully16}.}
\tablefoottext{c}{Weighted average value after LAB Survey of Galactic \ion{H}{i} \citet{kalberla05}.}
}
\end{table}

The radio emission from M\,83 was observed in a number of radio continuum studies that included polarimetry observations, which probe the magnetic fields. 
The early observations by \citet{sukumar87} and \citet{sukumar89} at 92\,cm, 20\,cm, and 6.3\,cm already traced ordered magnetic fields via radio polarised emission, 
which was enhanced outside of the optical spiral arms, forming the so called magnetic arms \citep[see Chapter~9 of][for a review]{beck19}. 
Strong Faraday depolarisation between 20\,cm, and 2.8\,cm was found by \citet{neininger93}. High-resolution sensitive observations 
with the Australian Telescope Compact Array (ATCA)\footnote{The Australia Telescope Compact Array is part of the Australia
Telescope National Facility, which is funded by the Commonwealth of
Australia for operation as a National Facility managed by CSIRO.} and the Very Large Array (VLA)\footnote{The Very Large Array is part of the National Radio Astronomy Observatory, which 
is a facility of the National Science Foundation operated under cooperative agreement by Associated Universities, Inc.} radio telescopes confirmed 
these findings \citep{frick16} and the morphology of the magnetic arms was compared to the spiral structures seen in the maps of the emission of warm dust, ionised gas, molecular gas, and 
atomic gas. The authors concluded that the material arms in some parts of the galaxy are closely followed by the magnetic arms while in the other a significant shift is visible.

Due to its brightness and angular extent, M\,83 has been observed many times with XMM-Newton \citep{jansen01} and Chandra \citep{weisskopf02} X-ray space observatories. 
The main scientific interest, however, focused on X-ray source populations \citep[e.g.][]{ducci13, long14} with only few publications 
analysing the emission from the hot X-ray gas. Especially relevant for this paper is the work by \citet{owen09}, who studied the diffuse 
X-ray emission from M\,83 and several other face-on galaxies. Nevertheless, that study used only 
one XMM-Newton observation (ObsID 110910201, see Table~\ref{xdat}) and aimed at global properties of the sample galaxies. In this paper, 
we take advantage of the combination of all available observations that allowed to obtain deep X-ray images, as well as very sensitive spectra of selected regions within the galactic disc.

To follow a similar analysis of the radio and X-ray data for NGC\,6946 \citep{wezgowiec16}, this paper presents the extended emission from the hot gas and the polarised radio emission of M\,83. The data were acquired by the XMM-Newton and the VLA telescopes. 
The parameters of the hot gas calculated from the model fits to the hot gas spectra from selected regions of the galaxy were compared to the properties of the magnetic fields, derived from the analysis of the 
radio emission.

We present the details of the data reduction and analysis in Sect.~\ref{obsred}, and the obtained results in Sect.~\ref{results}. 
Section~\ref{disc} includes the discussion of the results, that focuses mainly on correlations and comparisons
of the properties of the hot gas and of the magnetic fields of M\,83 with a special attention drawn to differences between 
the spiral arms and the interarm regions. We also compare our findings with the results of our previous study of NGC\,6946 \citep{wezgowiec16}.
We present our conclusions in Sect.~\ref{cons}.

\section{Observations and data reduction}
\label{obsred}

\subsection{Radio observations}
\label{radiored}

In this paper we re-analyse the VLA data already presented in \citet{Vogler05} and \citet{frick16}
to obtain the highest sensitivity to extended radio structures and consequently obtain information
about the strengths and energies of the total magnetic fields, as well as their regularities, in relevant areas of the spiral arm and interarm regions of M\,83. This was needed because \citet{frick16}
presented the VLA data at 4.86\,GHz that were merged with the single-dish data obtained with the 100-m Effelsberg radio telescope,
to recover the large-scale emission that was lost in the interferometer observations.
Nevertheless, we found that in the case of M\,83 the VLA observations at 4.86\,GHz include already majority of this emission and the larger extent of the emission presented in \citet{frick16}
results from the larger beam size used by these authors that increases the visibility of the large-scale emission detected by the Effelsberg telescope. 
Because for our work a smaller beam is desirable to better distinguish the radio emission from the spiral arms and the interarm regions, we decided
no to include the Effelsberg data. Furthermore, since no Effelsberg observations of M\,83 are available at 8.46\,GHz, such inclusion would lead to biases in both the Rotation Measure and the
Faraday depolarisation maps.

Polarisation radio continuum data were taken from the NRAO Science Data Archive\footnote{https://archive.nrao.edu/archive/advquery.jsp} 
and re-reduced by us, what allowed to obtain an optimal balance between the sensitivity to extended structures and the resolution, crucial 
for analysis of the properties of the magnetic fields in the spiral and the magnetic arms.
Observations were performed with the VLA within projects AS471 and AR431, in C and X bands respectively,
both in the compact CD-array configuration.

Default VLA frequency setups in both band were used, that is, two
sub-bands centred at 4.835 and 4.885\,GHz for C band, and 
at 8.435 and 8.485\,GHz for X band, respectively, each of sub-bands of 50\,MHz width.
Two separate pointings were observed in C band -- centred north-west and south-east from the 
galaxy centre. Only one pointing (centred at the galaxy core) was used in 
X band. Due to the large angular extent of M\,83, the size of the VLA primary beam allowed to cover 
at this frequency only the central part of the galaxy.

Each pointing was processed with the use of a standard data reduction routine in {\sc aips}\footnote{http://www.aips.nrao.edu/index.shtml}.
Raw correlator output was carefully inspected and all possible 
RFI or observing system errors were flagged. 
Phase calibration was used using the nearby point source J1313-333. The same
source was used for determination of polarisation leakage terms. Flux scale
and the polarisation position angle were established using the 3C286 calibrator.
The relatively bright core of M\,83 allowed to effectively determine phase
corrections -- four loops of self-calibration that determined phase-only corrections
were followed by two cycles with phase-and-amplitude corrections. 

The final maps were calculated using the {\sc imagr} task with the {\sc ROBUST}
parameter set to 2 and 3 for C and X band data, respectively.
All Stokes I, Q, and U maps were convolved to a common circular Gaussian of
15$\arcsec$ half power beamwidth (HPBW). 
Maps of both pointings in C band were combined using the task {\sc flatn}, 
prior to combination of Stokes Q and U maps into polarised
intensity and polarisation angle maps. 
Final maps of total and polarised intensities 
have r.m.s. noise levels of about 30 and 25\,$\mu$Jy/beam in C band, and
30 and 15\,$\mu$Jy/beam in X band, respectively.

\subsection{X-ray observations}
\label{xrayred}

\begin{table*}[ht]
        \caption{\label{xdat}Characteristics of the X-ray observations of M\,83}
\centering
\begin{tabular}{lccccc}
\hline\hline
Obs ID	  	&EPIC-pn filter&EPIC-pn observing mode	&total time [ks] & clean time [ks]\\
\hline
110910201	& thin	       & extended full frame	& 27.0		 & 18.6		  \\
503230101	& medium       & extended full frame	& 30.6		 & 13.3		  \\
552080101	& medium       & extended full frame	& 25.6		 & 22.4		  \\
{\bf 723450101}	& thin	       & full frame	 	& 50.0		 & 50.0		  \\
723450201	& thin	       & full frame	 	& 55.0		 & 36.2		  \\
729561001	& thin	       & full frame	 	& 29.8		 & 16.3		  \\
{\bf 729561201}	& thin	       & full frame	 	& 26.8		 & 26.8		  \\
{\bf 761620101}	& thin	       & full frame	 	& 60.0	 	 & 59.4		  \\
{\bf 761620201}	& thin	       & full frame	 	& 72.1		 & 35.9		  \\
\hline
\end{tabular}
	\tablefoot{
Observations that were used in the spectral analysis are marked with boldface.
}
\end{table*}

M\,83 has been observed nine times between 2003 and 2016 with the XMM-Newton telescope.
Table~\ref{xdat} presents the details of all observations. All of them were processed using the SAS 17.0.0 package \citep{gabriel04} and its 
standard reduction procedures. The routine of tasks {\sc epchain} and {\sc emchain} produced for each observation event lists for two EPIC-MOS cameras \citep{turner01} 
and the EPIC-pn camera \citep{strueder01}. The event lists were filtered for periods of intense radiation of high-energy background, using the good time interval (GTI) tables 
based on the derived light curves of high-energy emission. The standard thresholds of 0.35\,cts/s and 0.4\,cts/s for the EPIC-MOS and the EPIC-pn cameras, respectively, were used, to 
assure good quality data of the diffuse soft emission. 
The resulting lists were checked for the residual existence of soft proton flare contamination, which
could influence the faint extended emission. To do that, we used a script\footnote{http://xmm2.esac.esa.int/external/xmm\_sw\_cal/\\background/epic\_scripts.shtml\#flare}
that performs calculations developed by \citet{deluca04}.
We found that observations 110910201, 723450201, and 729561001 were significantly contaminated by soft proton radiation. Because the more conservative approach of filtering the data 
(lower threshold limits in the creation of GTI tables) did not reduce the contamination, we decided to exclude from the spectral analysis these three observations, as well as the two medium filter observations, ObsIDs 503230101 and 552080101, in which the galaxy was in an off-axis position (close to the FOV edge). All data was, however, used to produce deep images of the distribution of the X-ray emission from M\,83.
To ensure the best-quality data (crucial to analyse diffuse emission), we only used events with FLAG=0 and PATTERN$\leq$4 (EPIC-pn) or FLAG=0 and PATTERN$\leq$12 (EPIC-MOS) in the following data
processing.

The images, background images, and exposure maps (without vignetting correction) were produced separately for each observation using the filtered event lists and the 
images script\footnote{http://xmm.esac.esa.int/external/xmm\_science/\\gallery/utils/images.shtml}, modified by the authors
to allow adaptive smoothing.
All images were then combined into two final EPIC images, soft and medium in the 0.2-1 and 1-2\,keV energy bands, respectively. The resulting images 
were adaptively smoothed with a maximum smoothing scale of 30$\arcsec$ full width at half maximum (FWHM) and a signal to noise ratio of 30, to efficiently detect structures of the diffuse emission.
The r.m.s values were obtained by averaging the emission over a large source-free areas in the final maps.

The same procedure of image creation was repeated with the use of event lists from which all point sources within the D$_{25}$ diameter of M\,83 were excluded (see below for details).
This was done using the same approach as in the creation of the re-filled blank sky background maps - 
ghostholes\_ind\footnote{ftp://xmm.esac.esa.int/pub/ccf/constituents/extras/\\background/epic/blank\_sky/scripts}.
Such treatment of the data allowed to obtain maps of diffuse emission in which all regions of excluded point sources are filled with emission close to extracted regions by sampling adjacent events and
randomising spatial coordinates\footnote{http://xmm2.esac.esa.int/external/xmm\_sw\_cal/\\background/blank\_sky.shtml\#BGsoft}. 
Although usually used to prepare background maps, the procedure proved to be appropriate also for the real source data.
Section~\ref{dist} presents maps of soft (0.2-1\,keV) and medium (1-2\,keV) X-ray emission, together with a corresponding
hardness ratio map, defined as $$HR=\frac{{\rm med}-{\rm soft}}{{\rm med}+{\rm soft}},$$ for images with and without detected point sources.

For a better comparison of the X-ray and radio maps of M\,83, the X-ray event lists without detected point sources were once again used to produce 
an image that includes the entire diffuse emission, that is, in the band 0.2-2\,keV. Due to the absence of point sources, a convolution of this image with a Gaussian profile resulted 
in a constant resolution of 15\arcsec and allowed a direct comparison of the distribution 
of X-ray and radio emission.

The spectral analysis, as mentioned above, was performed using only the four observations that provided the X-ray data uncontaminated by soft protons and pointed at the centre of M83. 
To create spectra we only used the event lists from the EPIC-pn camera because it offers the highest sensitivity in the soft energy band. 
Only the emission above 0.3\,keV was analysed because the internal noise of the pn camera is too high below this
limit\footnote{http://xmm.esac.esa.int/external/xmm\_user\_support/\\documentation/uhb}. Although this is not crucial when combined with MOS cameras to produce images, it is important
to exclude the softest emission below 0.3\,keV to obtain reliable good-quality spectra.

To search for point sources within the D$_{25}$ disc of M\,83 we used the {\it edetect\_stack} meta task, that maximises the sensitivity of 
the source detection using the data combined from all observations. This procedure provided a list of positions of the detected point sources. 
The area was individually chosen for each source to ensure that all pixels brighter than the surrounding background are excluded.
These areas were then used to construct regions of diffuse X-ray emission free from significant contribution from point sources. 
The non-default procedure of exclusion of the detected point sources helped to keep more diffuse emission in the final spectra.
However, because we expected some contribution from the PSF wings, we added a power-law component to our models to account for any residual emission.
This power-law component was also needed to account for unresolved point sources.
The background spectra were obtained using blank sky event lists \citep[see][]{carter07}.
The blank sky event lists were filtered in the same manner as the source event lists.
For each spectrum we produced response matrices and effective area files. For the latter, detector maps needed for extended
emission analysis were also created. Next, for each region the spectra, acquired for each of the four observations separately, were merged into a final spectrum that included the information 
from the response matrices and ancillary files, as well as the background emission. This was done with the use of the SAS task {\it epicspeccombine}. The advantage of this approach, when compared to simultaneous fitting 
of all spectra of a given region, is that the source spectra and the background spectra are merged separately and only after that the background subtraction is performed. This leads to better 
results due to higher signal-to-noise ratio of the merged spectra, comparing to lower signal-to-noise of individual spectra. Finally, the spectra 
of each of the studied region were binned, to further improve the signal-to-noise ratio. 
To obtain a reasonable number of bins at the same time, we chose to have 25 total counts per energy bin. The spectra were fitted using XSPEC~12 \citep{arnaud96}.

For the overlays we also used the XMM-Newton Optical Monitor data acquired during the same observations and produced an image in the UVW1 filter, using the standard SAS {\sc omchain} procedure and 
the {\sc emosaic} task to combine the individual images.

\section{Results}
\label{results}

In Sect.~\ref{radio} we present the radio maps at 4.86\,GHz and 8.46\,GHz, both in total and polarised intensity, as well as
the Rotation Measure and the Faraday depolarisation maps that provide information about the properties of the halo of M\,83. 
The sensitivities reached in the intensity maps are presented in Table~\ref{radiosens}.

Section~\ref{dist} presents the maps of the distribution of the
diffuse X-ray emission from the hot gas in M\,83 and its spectral analysis is presented in Sect.~\ref{spectra}.

\subsection{Radio emission}
\label{radio}

\begin{figure*}[ht]
\resizebox{0.495\hsize}{!}{\includegraphics[clip]{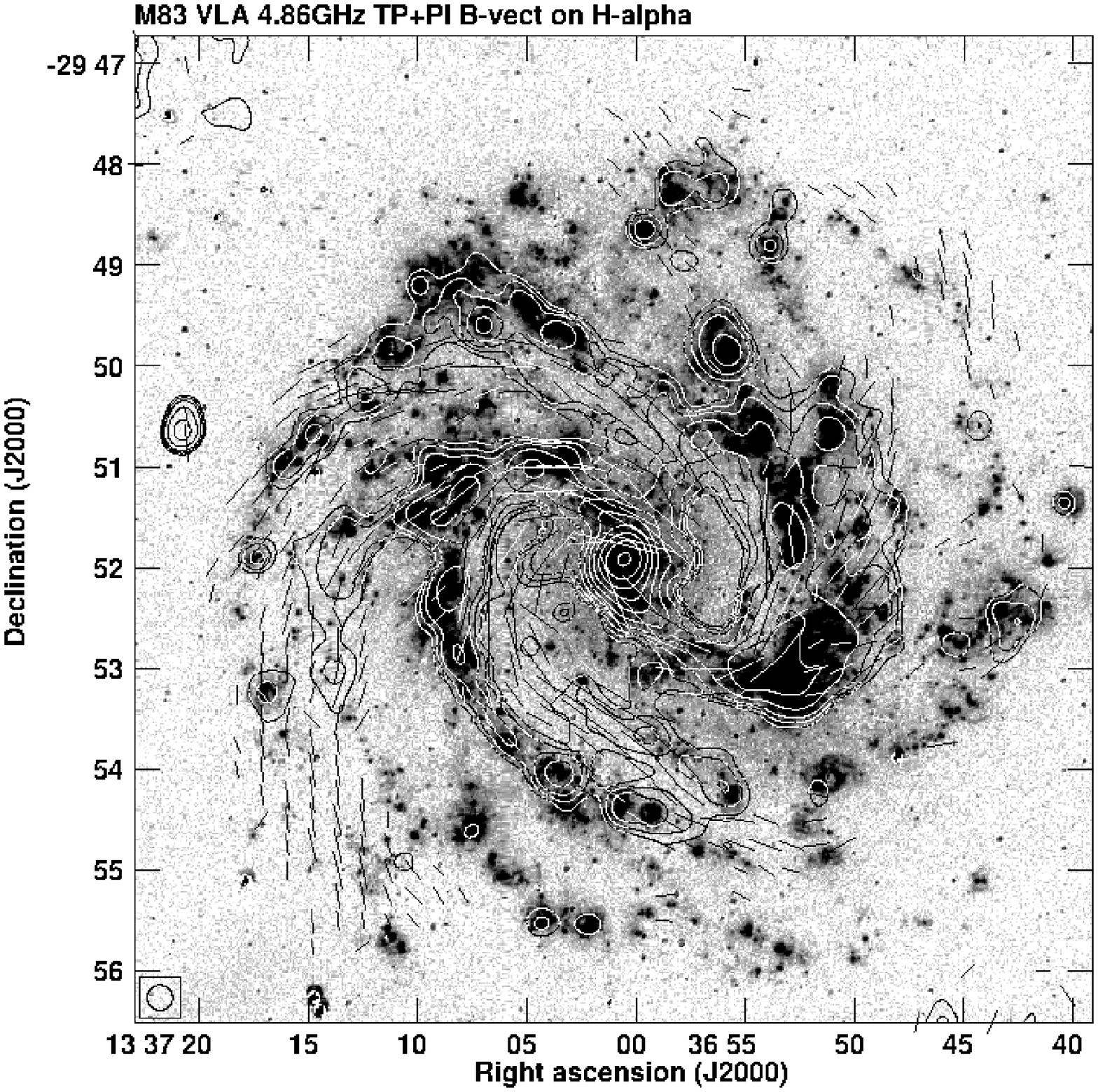}}
\resizebox{0.505\hsize}{!}{\includegraphics[clip]{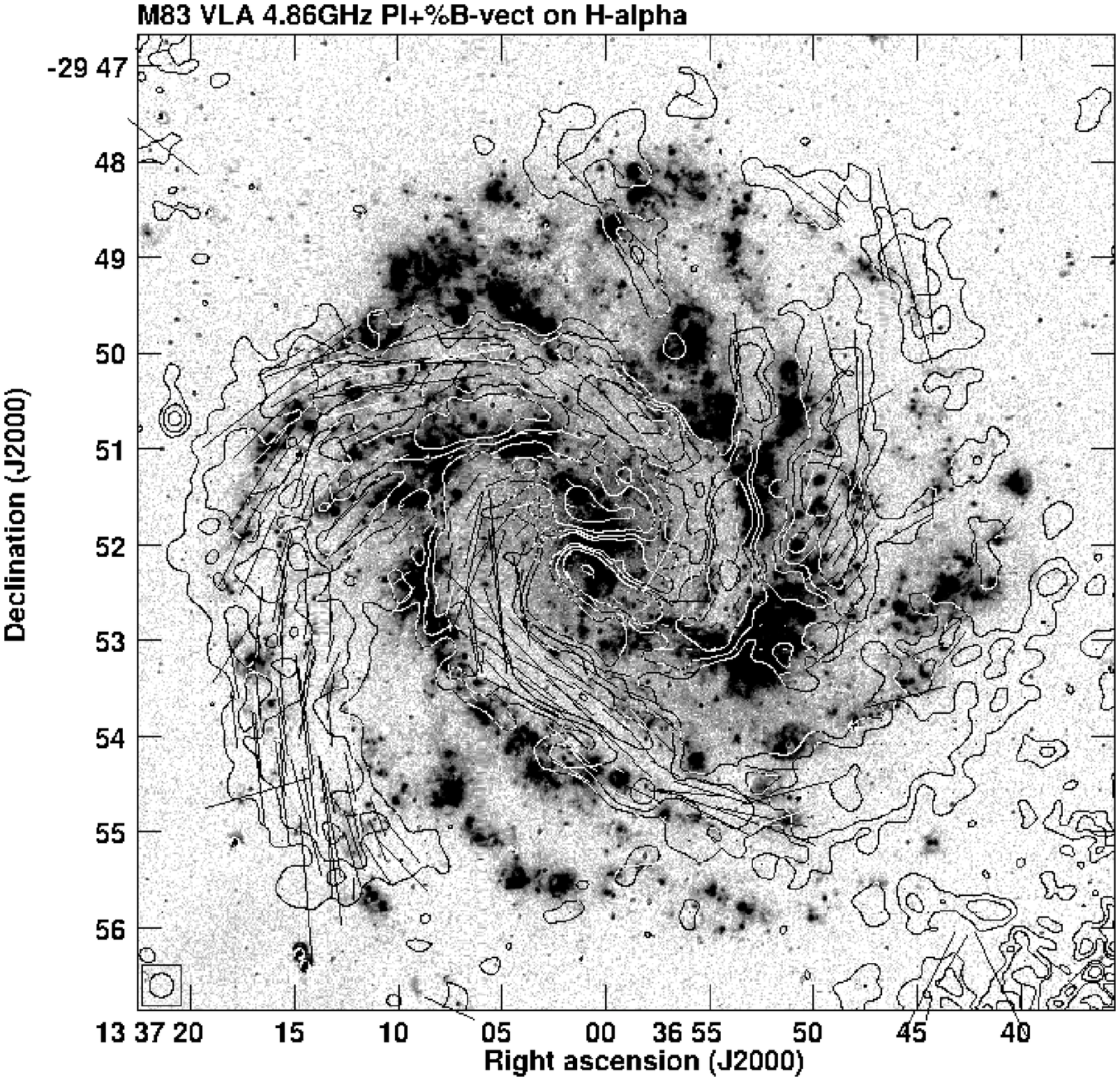}}
\caption{
	Map of total (left) and polarised (right) radio intensity at 4.86\,GHz 
	($\lambda$ 6\,cm) of M\,83 overlaid on the H$\alpha$ map. The vectors show the orientation of the magnetic fields 
	and their lengths are proportional to the polarised intensity (left) or degree of polarisation (right). 
	The angular resolution is 15$\arcsec$.
        }
\label{6radio}
\end{figure*}

The total radio emission at 4.86\,GHz closely follows the spiral arms of M\,83, while the 
polarised intensity is mostly confined to the interarm regions (left and right panels of Fig.~\ref{6radio}, respectively), forming the magnetic arms. 
The most prominent is the narrow arm that starts at the eastern end of the galactic bar 
and avoiding the star-forming regions turns to the north-west, already outside of the main body of the galaxy. 
Another large patch of the polarised emission is visible north of the south-western end of the bar, together with some patchy emission to the west. 
Finally, a broad magnetic arm is visible in the eastern part of the disc, 
though several clumps of the H$\alpha$ emission are also present in this area. 
The highest degree of polarisation is visible in the north-east and the south-west of the galactic disc. 

The same features are visible in the combined VLA+Effelsberg map shown by \citet{frick16} with the emission appearing slightly more extended, especially in the northern and north-western part of the disc. However, as mentioned in Sect.~\ref{intro}, this is 
an effect of a larger beam (of 22\arcsec compared to 15\arcsec of our map) and not the recovered emission, lost in the interferometric observations. 
The patchy emission visible at this position in our map would become homogenous when smoothed to a resolution of 22\arcsec. This resolution, however, would make the separation of the spiral arm and the interarm emission more difficult.

\begin{figure*}[ht]
\resizebox{0.5\hsize}{!}{\includegraphics[clip]{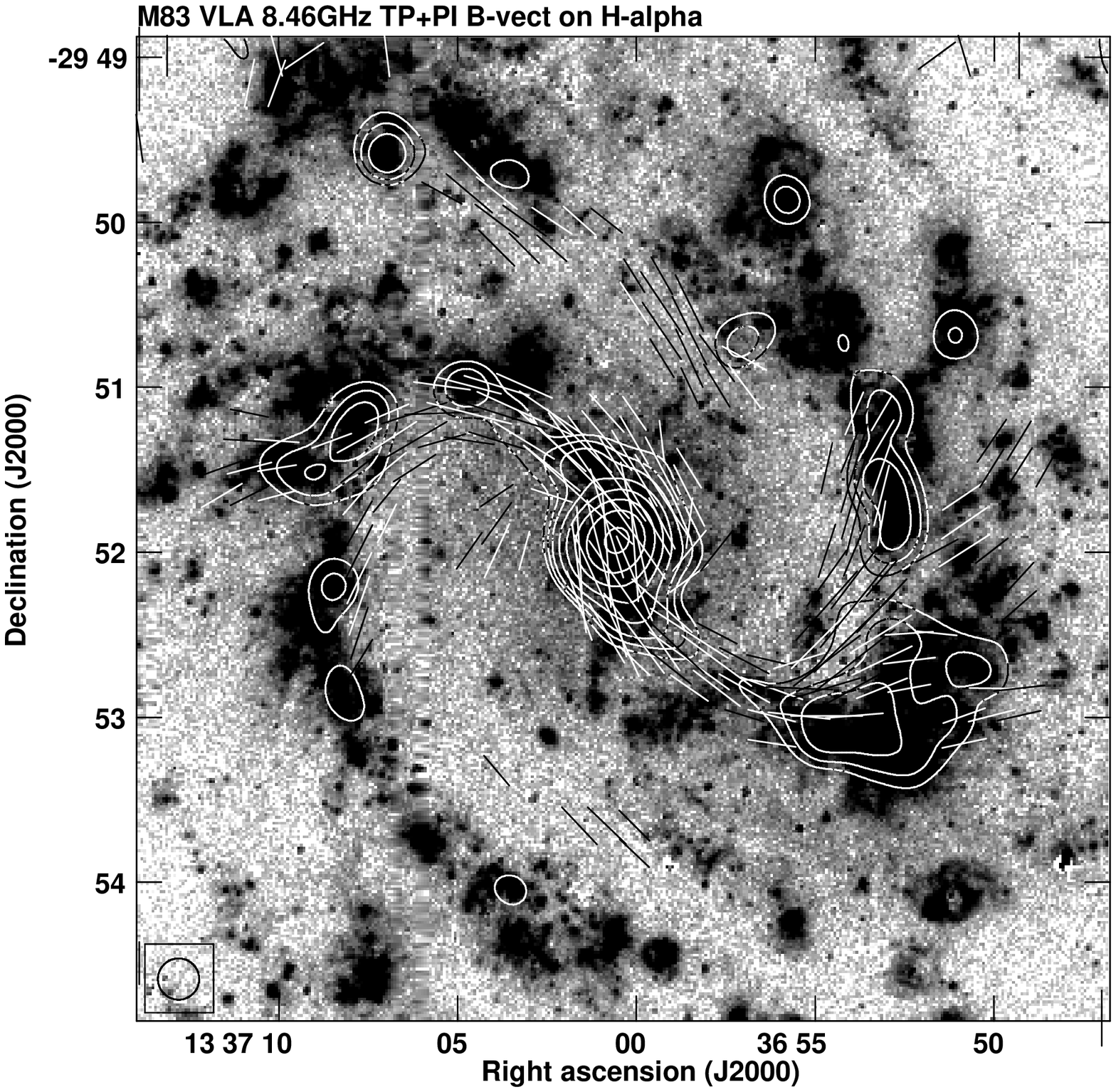}}
\resizebox{0.5\hsize}{!}{\includegraphics[clip]{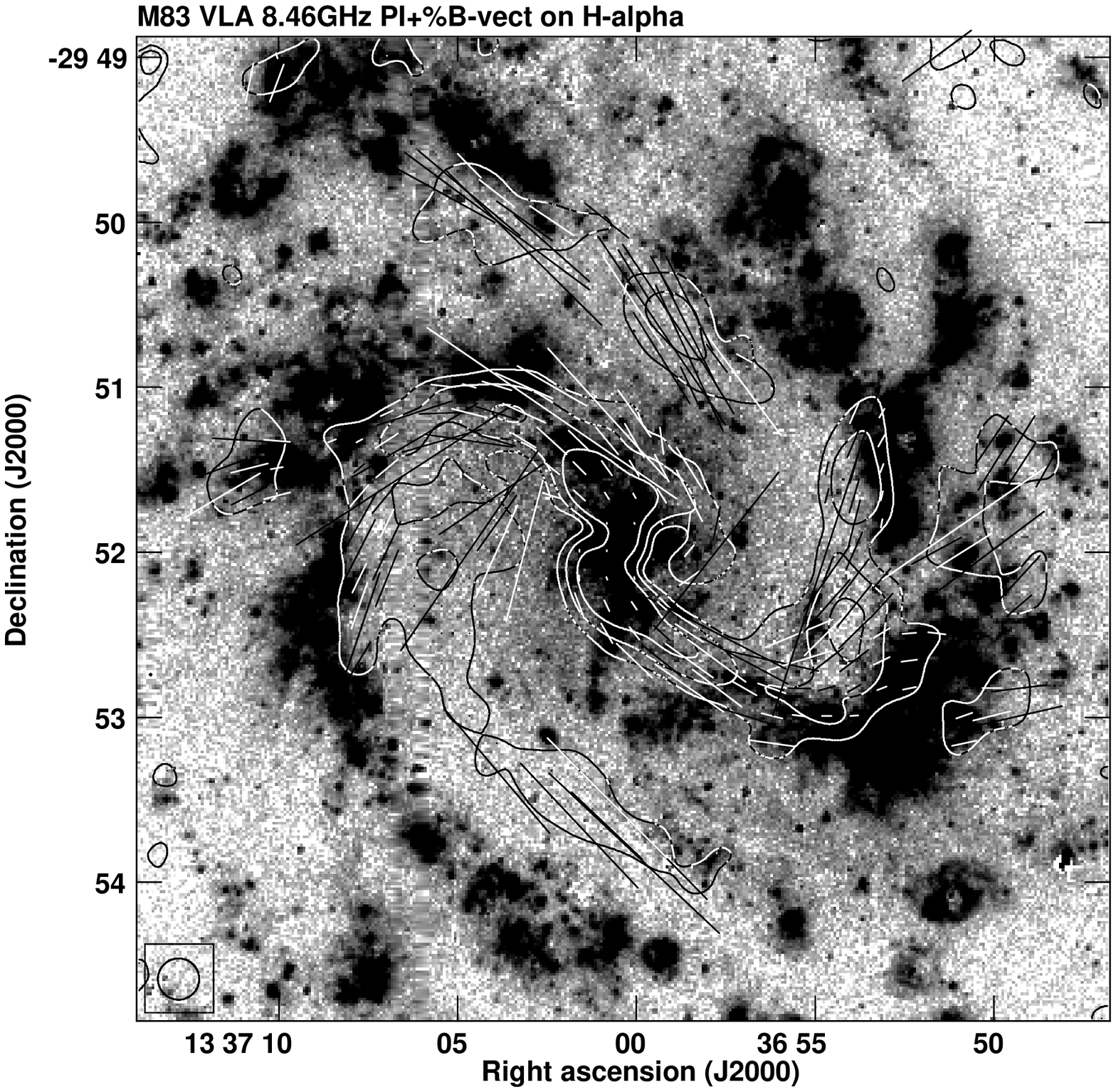}}
\caption{
        Map of total (left) and polarised (right) radio intensity at 8.46\,GHz
        ($\lambda$ 3.6\,cm) of M\,83 overlaid on the H$\alpha$ map. The vectors show the orientation of the magnetic fields 
        and their lengths are proportional to the polarised intensity (left) or degree of polarisation (right). 
        The angular resolution is 15$\arcsec$.
        }
\label{3radio}
\end{figure*}

At 8.46\,GHz only the bright central parts of M\,83 are visible. The total intensity map (left panel of Fig.~\ref{3radio}) 
shows the emission along the bar with a bright core region. In the polarised intensity map (right panel of Fig.~\ref{3radio}) the inner part of the south-western magnetic arm is still visible. 

\begin{figure*}[ht]
\resizebox{0.51\hsize}{!}{\includegraphics[clip]{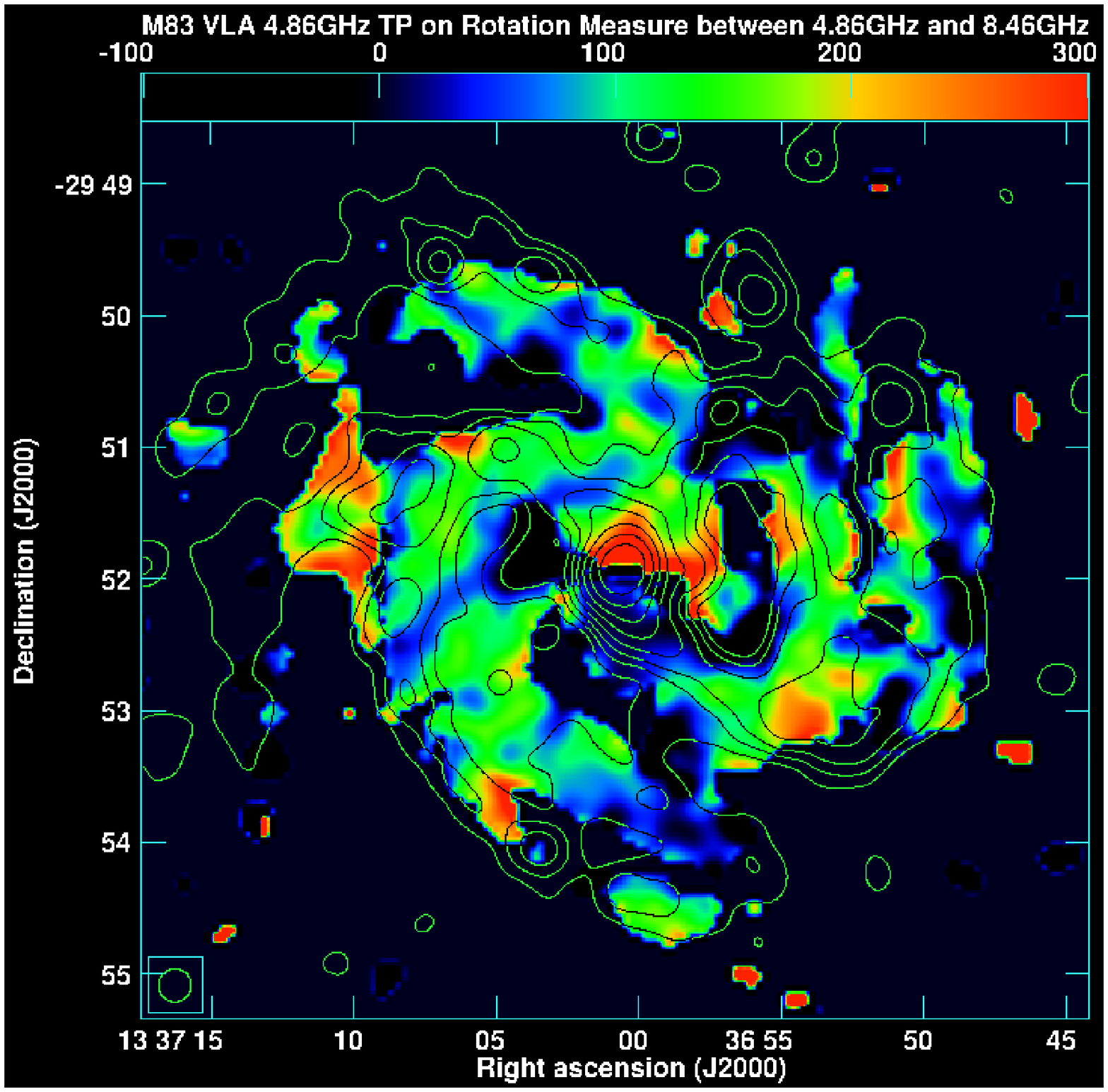}}
\resizebox{0.49\hsize}{!}{\includegraphics[clip]{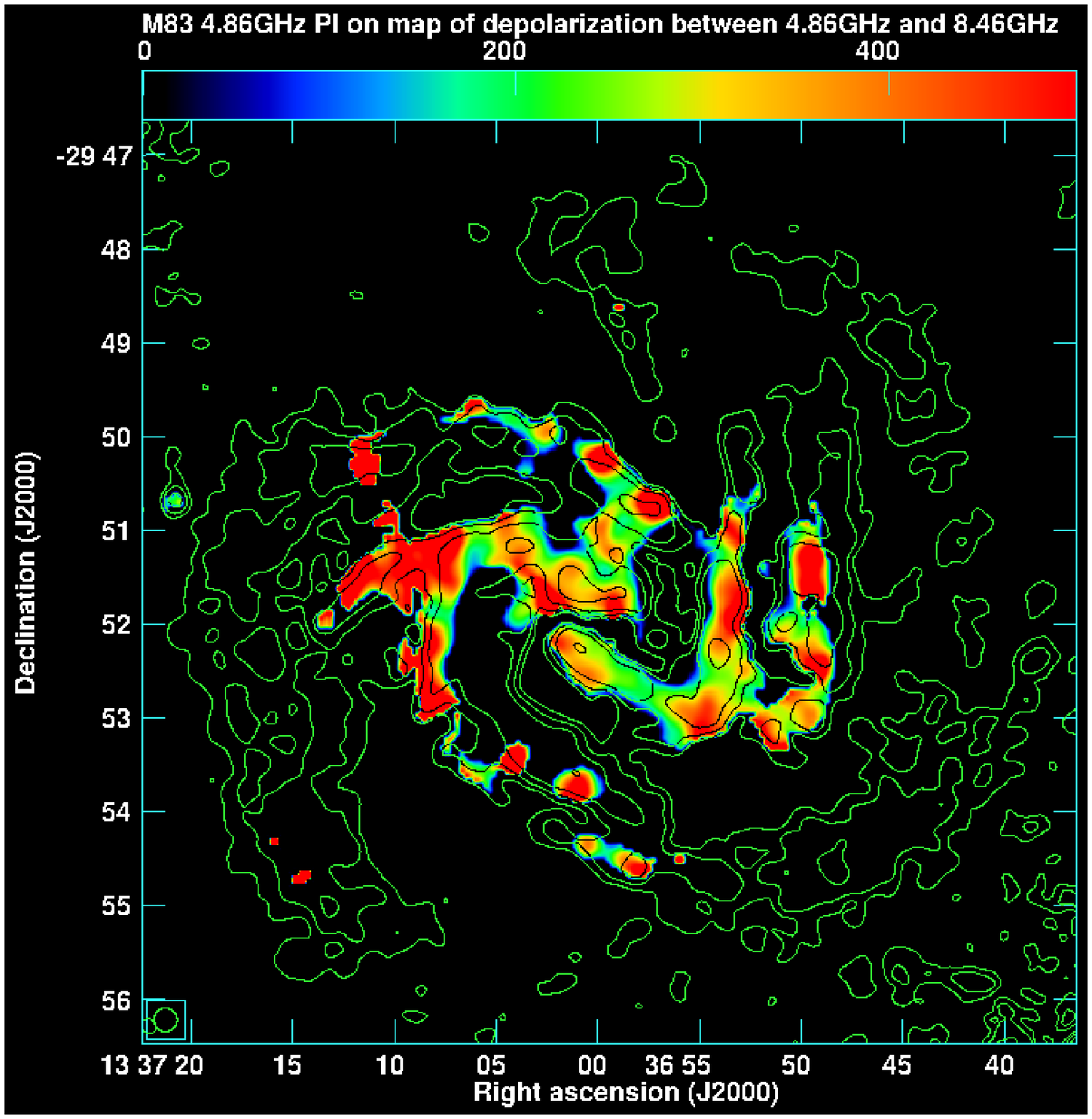}}
\caption{
	Left: Map of Rotation Measure between 4.86 and 8.46\,GHz overlaid with contours of total radio intensity at 
	4.86\,GHz.
	Right: Map of Faraday depolarisation between 4.86 and 8.46\,GHz overlaid with contours of polarised radio intensity at
	4.86\,GHz. The angular resolution of both maps is 15$\arcsec$.
	}
\label{m83rm}
\end{figure*}

To obtain information about the regular magnetic field in the line-of-sight, that is, the halo component, we produced a Rotation Measure map (left panel of Fig.~\ref{m83rm}),
using the polarisation angle maps at both frequencies. Next, with the use of the maps of degree of polarisation we produced a Faraday depolarisation map (right panel of Fig.~\ref{m83rm}). 

In the Rotation Measure map areas of uniform values (100-120\,rad/m$^2$) are visible along the magnetic arms, which confirms
the existence of large-scale regular magnetic fields. The Faraday depolarisation map is limited to the central parts of the disc. 
The value of depolarisation is mostly below 0.5, locally reaching 0.16 (lower values mean stronger depolarisation). 

\begin{table}[ht] 
        \caption{\label{radiosens}Sensitivities of the radio maps of M\,83.}
\centering
                \begin{tabular}{lrr}
\hline\hline
			&	4.86\,GHz	&	8.46\,GHz	\\
\hline
total intensity 	&	72\,$\mu$Jy	&	206\,$\mu$Jy	\\			
polarised intensity 	& 	25\,$\mu$Jy	&	39\,$\mu$Jy	\\
\hline
\end{tabular}
\end{table}

\subsection{Distribution of the X-ray emission}
\label{dist}

In this paper we focus only on the analysis of the distribution and the spectral properties of the diffuse X-ray emission from M\,83. The point source 
populations in this galaxy are beyond the scope of our work and their detailed analysis can be found in \citet{ducci13}. 

\begin{figure*}[ht]
                        \resizebox{0.5\hsize}{!}{\includegraphics{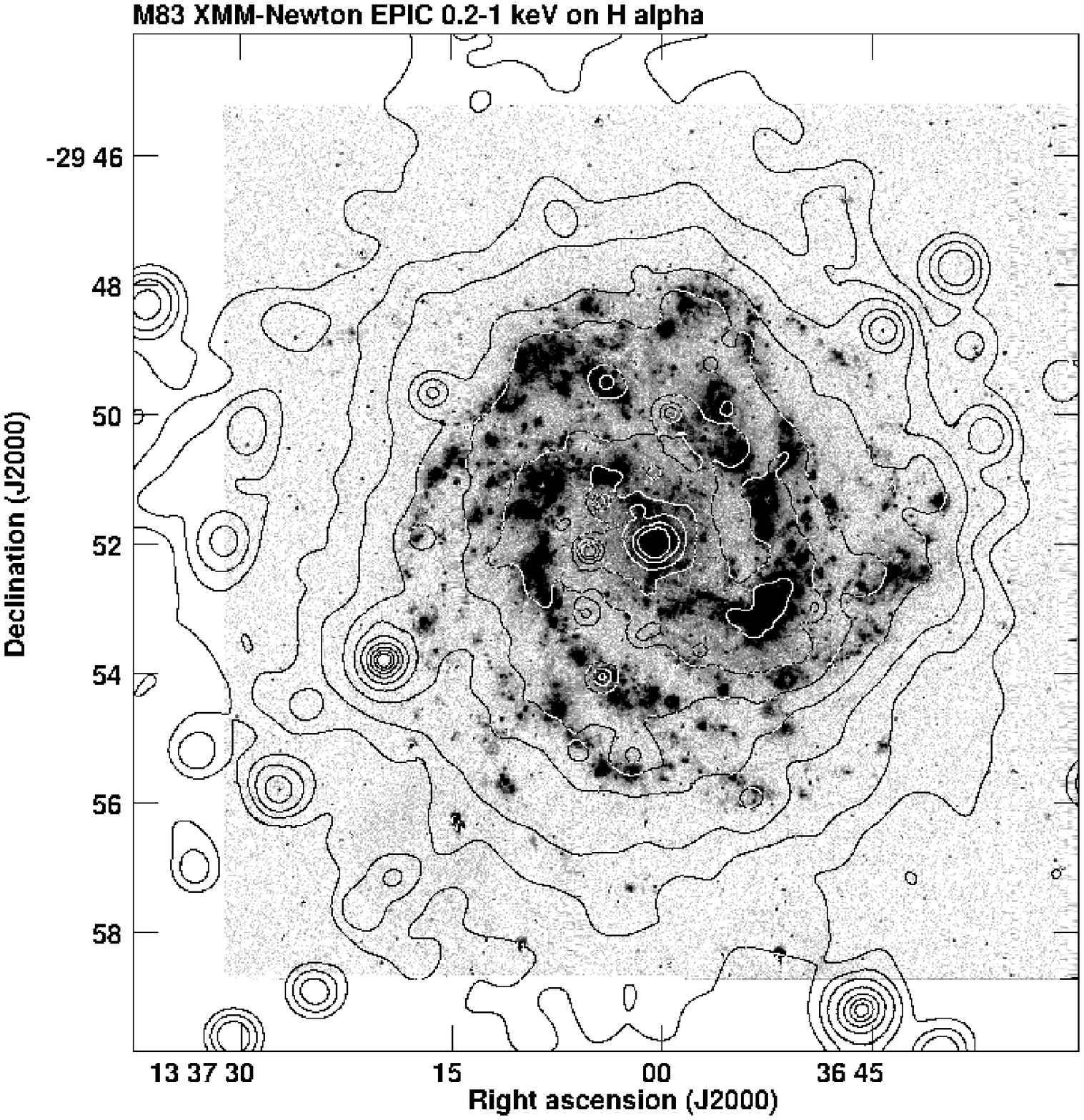}}
                        \resizebox{0.5\hsize}{!}{\includegraphics{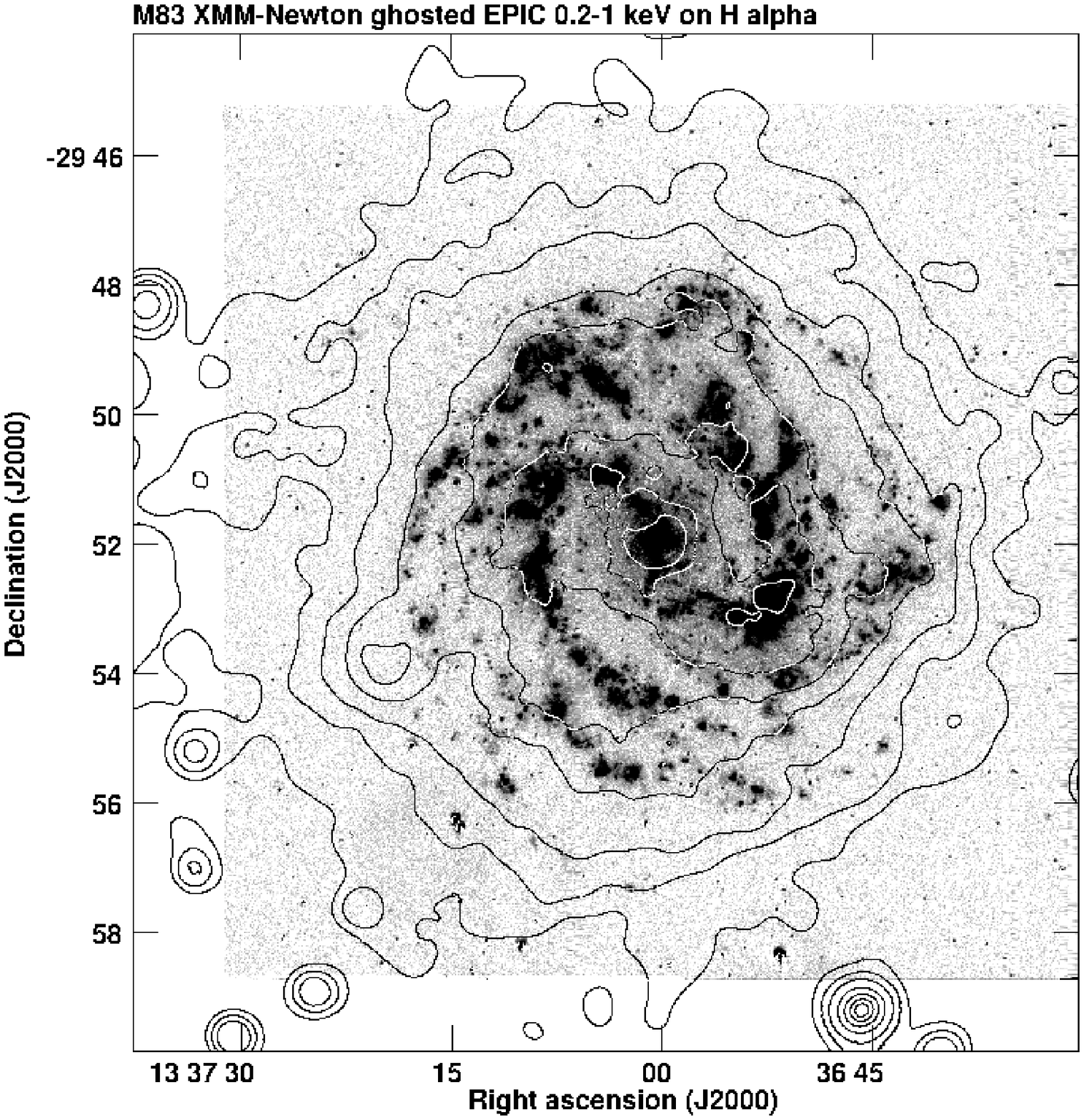}}
                \caption{
                Left: Map of soft X-ray emission from M\,83 in the 0.2 - 1 keV band overlaid onto an H$\alpha$ image. The
                contours are 3, 5, 8, 16, 32, 64, 128, 256, 512, and 1024 $\times$ rms. The map is adaptively smoothed with the
                largest scale of 30$\arcsec$. Right: Same map, but with point sources excluded from the galactic disc (see text for details).
                }
                \label{m83xsoft}
        \end{figure*}

The soft X-ray emission from M\,83 is visible across the entire star-forming disc (Fig.~\ref{m83xsoft}). The brightest central parts only slightly resemble 
a spiral structure and the exclusion of several point sources from the map does not reveal significant changes (right panel of Fig.~\ref{m83xsoft}). Nevertheless, the emission from the 
diffuse X-ray gas can be visible to larger distances (of the order of 3\arcmin or around 4\,kpc) from the star-forming disc. 

\begin{figure*}[ht]
                        \resizebox{0.5\hsize}{!}{\includegraphics{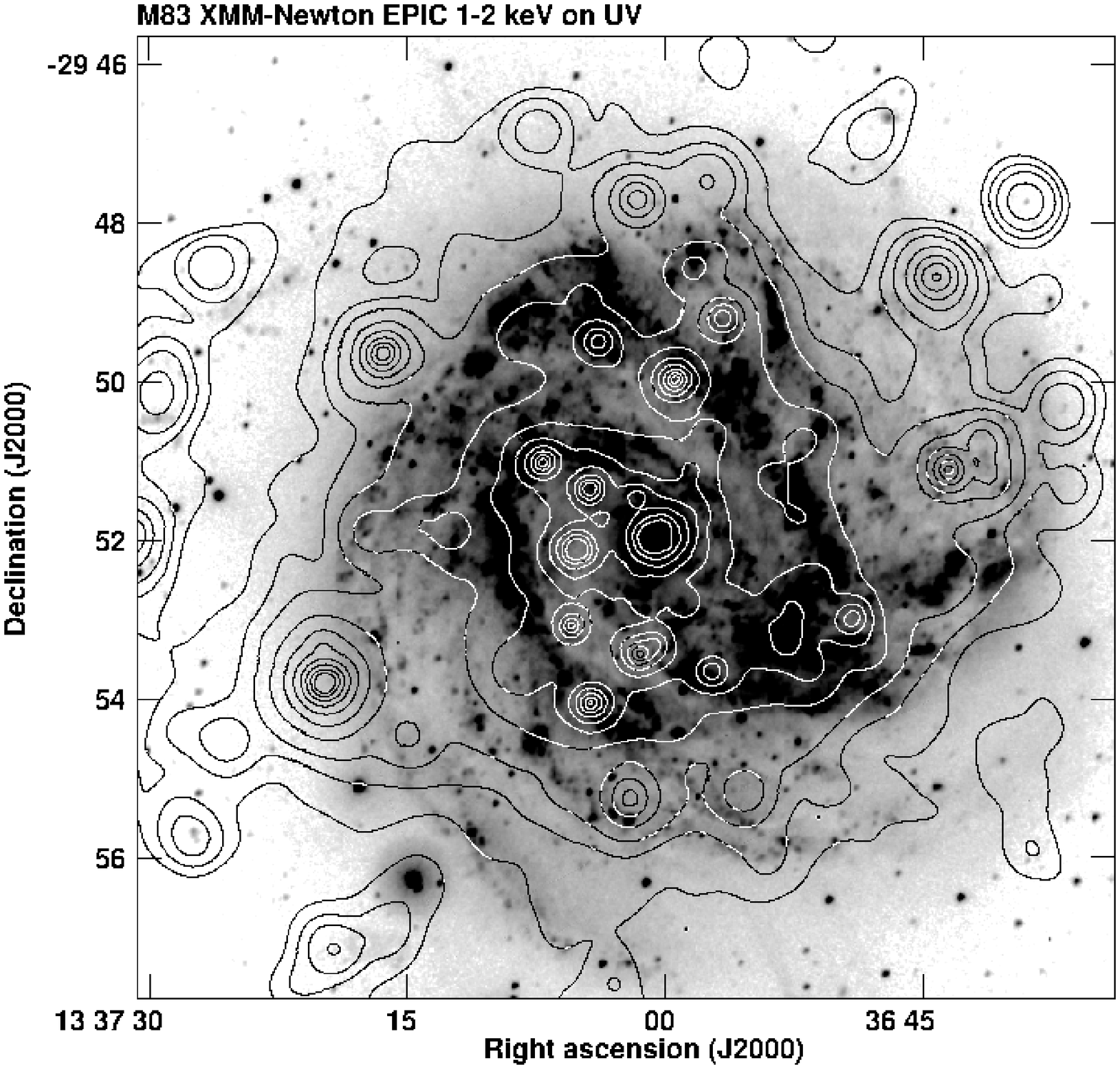}}
                        \resizebox{0.5\hsize}{!}{\includegraphics{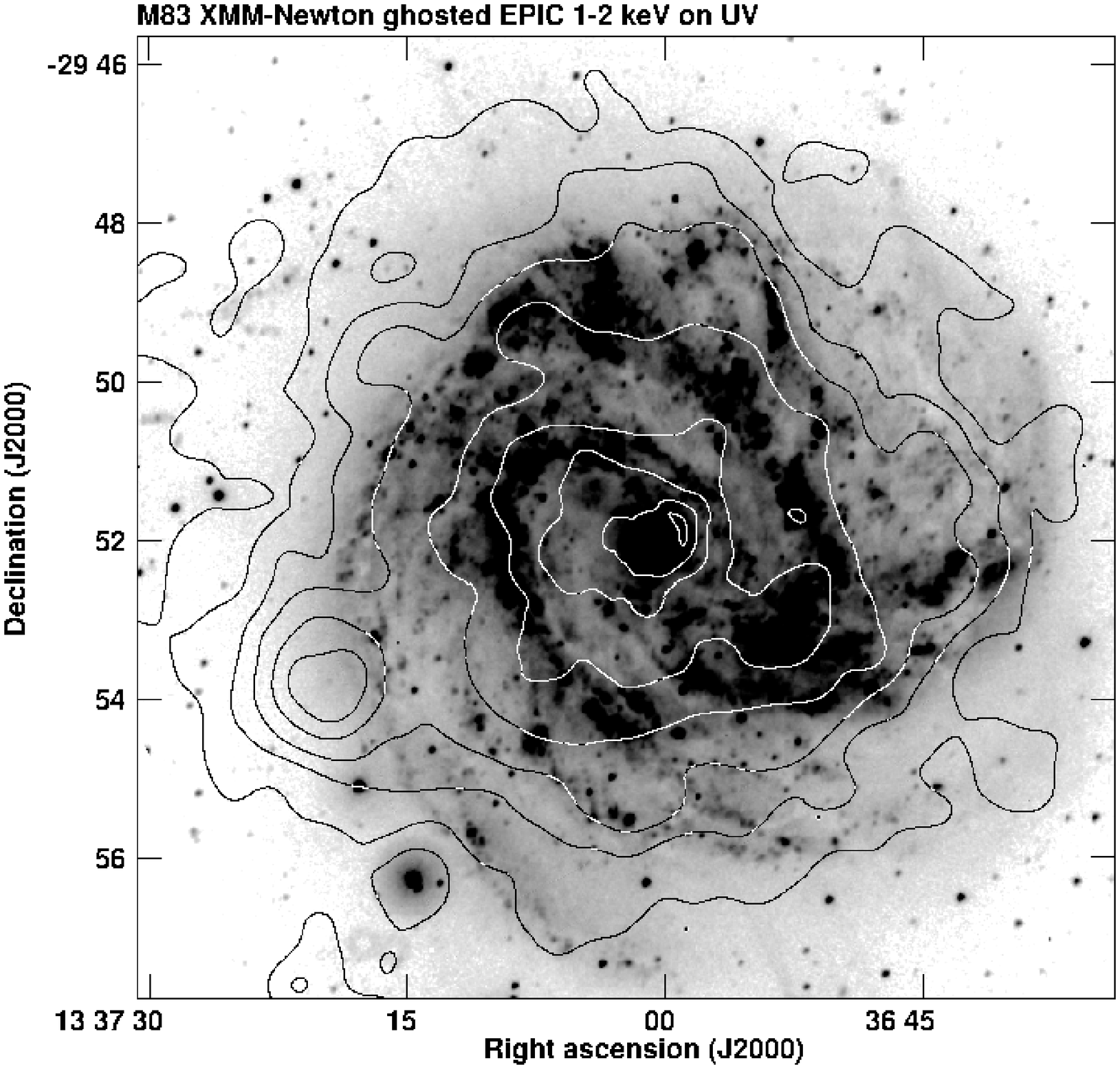}}
                \caption{
                Left: Map of medium X-ray emission from M\,83 in the 1 - 2 keV band overlaid on the XMM-Newton Optical Monitor UVW1 filter image.
                The contours are 3, 5, 8, 16, 32, 64, 128, 256, 512, and 1024 $\times$ rms. The map is adaptively smoothed
                with the largest scale of 30$\arcsec$. Right: Same map, but with point sources excluded from the galactic disc (see text for details).
                }
                \label{m83xmedium}
        \end{figure*}

The map of the harder X-ray emission reveals a higher number of point sources with the extent of the diffuse emission closely following the star-forming disc (Fig.~\ref{m83xmedium}). 
No significant extensions beyond this disc are visible. In the central part of the disc, an area of enhanced X-ray emission is visible south-west from the centre, clearly associated with 
an area of intense star formation (right panel of Fig.~\ref{m83xmedium}). Also here, no clear spiral pattern of the X-ray emission is visible.

To check the relative intensity of the soft (0.2-1\,keV) and medium (1-2\,keV) energy bands, we produced the hardness ratio maps, with and without the point sources. Especially the `ghosted'
map shows that the X-ray emission is rather uniform throughout the disc and apart from the slightly harder emission (hardness ratio of around -0.1) 
from its central parts, an area of softer emission (HR of around -0.6) is visible at the position of the bright star-forming region in the western spiral arm. 

\begin{figure*}[ht]
                        \resizebox{0.5\hsize}{!}{\includegraphics[clip]{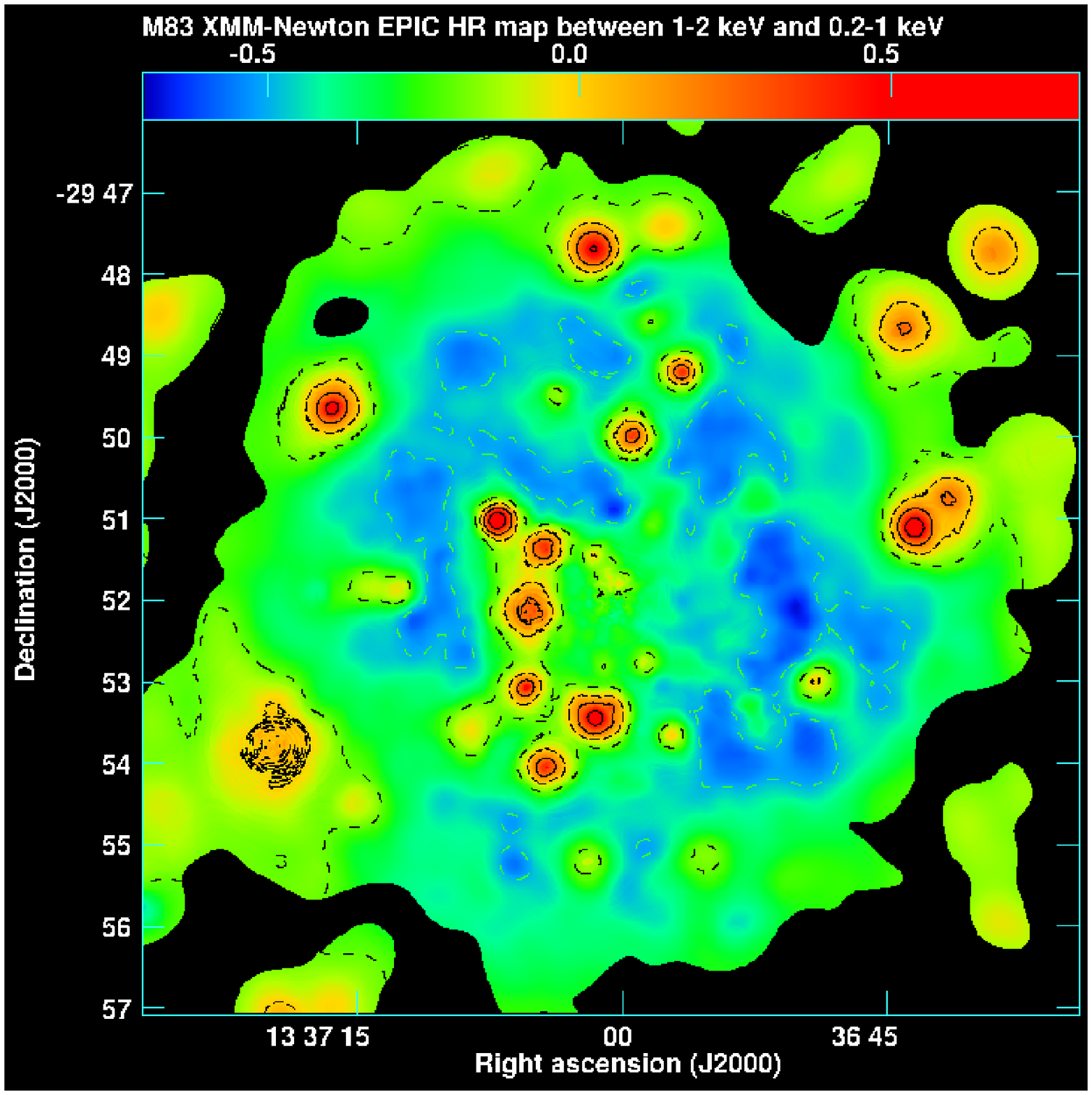}}
                        \resizebox{0.495\hsize}{!}{\includegraphics[clip]{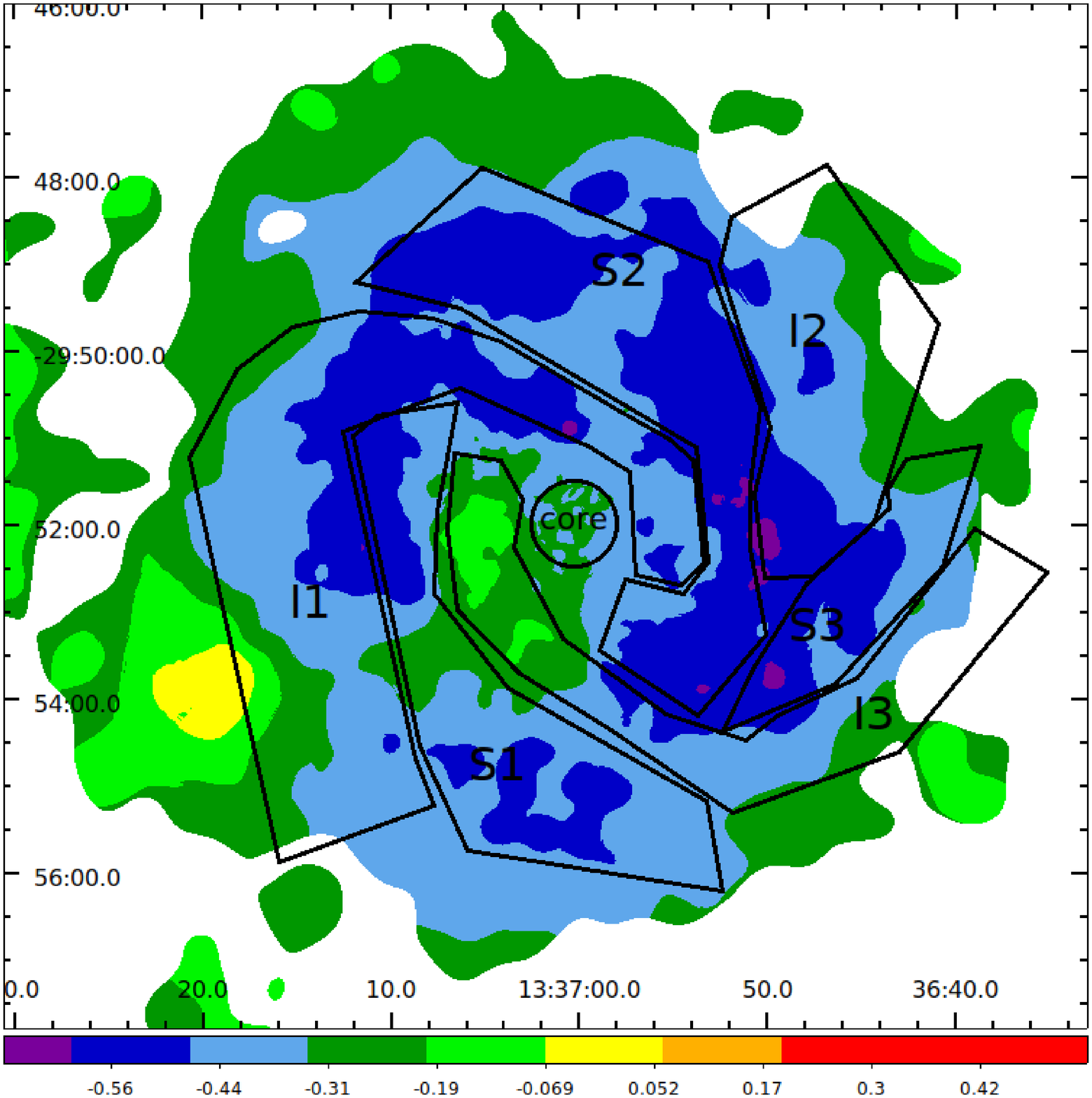}}
                \caption{Maps of the hardness ratio between medium and soft X-ray emission from M\,83 (Figs.~\ref{m83xmedium}
                         and~\ref{m83xsoft}). Both maps are truncated at the 3$\sigma$ level of the 1-2\,keV map. Left: Map with point sources and contours of -0.5, -0.2, 0, 0.2, and 0.5.
                        Right: Map without point sources overlaid with regions used for spectral analysis.}
                \label{m83hr}
        \end{figure*}

\subsection{Spectral analysis of the X-ray emission}
\label{spectra}

\begin{figure}[ht]
\resizebox{\hsize}{!}{\includegraphics[clip]{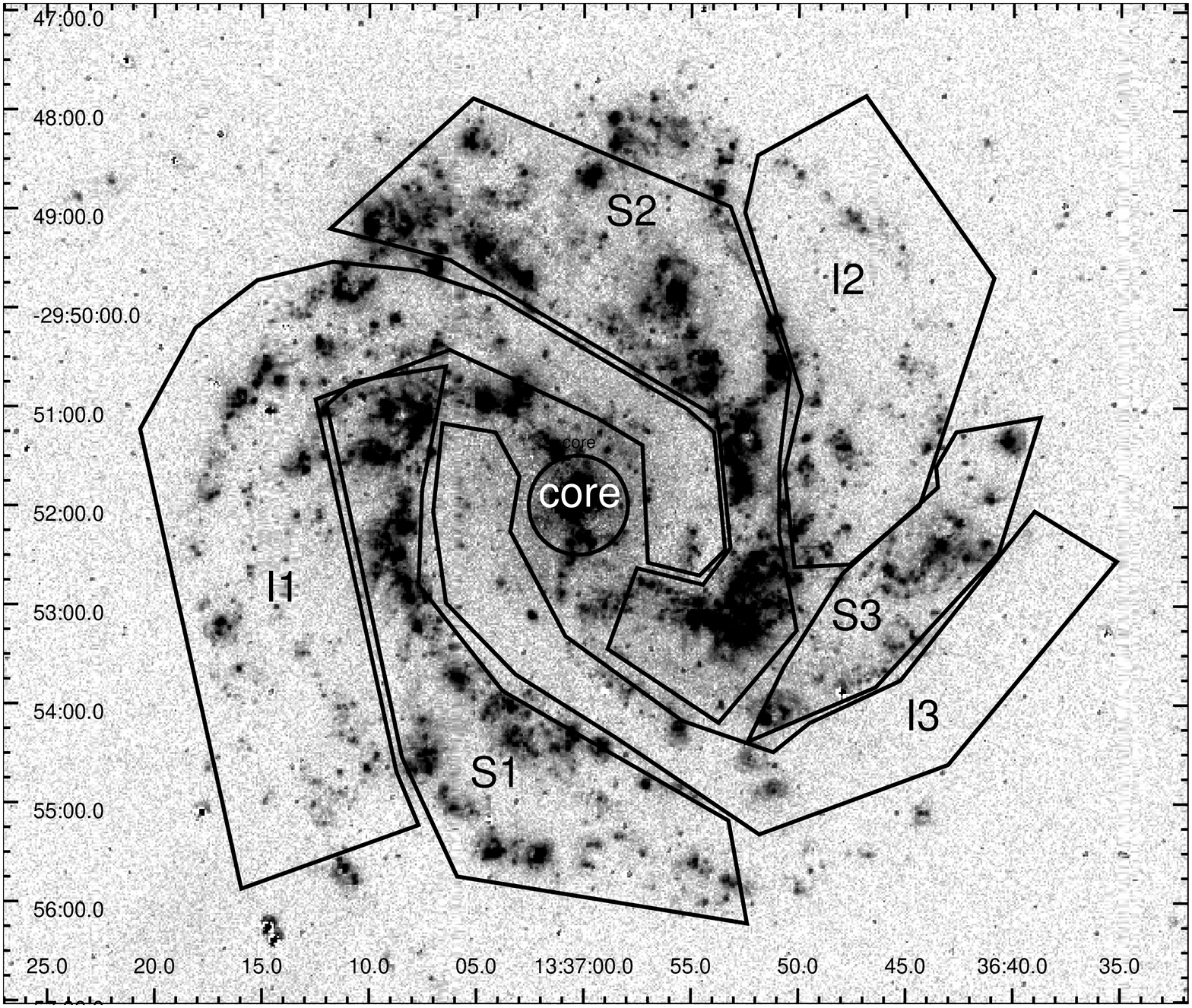}}
        \caption{Regions of diffuse X-ray emission in M\,83 (see text for a detailed description) overlaid on the H$\alpha$ map.
        }
\label{m83xregs}
\end{figure}

To study the properties of the hot gas of M\,83 we performed the spectral analysis of the diffuse X-ray 
emission. The polarised radio map at 4.86\,GHz (right panel of Fig.~\ref{6radio}), together with the H$\alpha$ map,
was used to identify regions of the spiral arms and the interarm space in M\,83. 
This allowed to construct spectral areas separately for the star-forming regions of the spiral arms and 
for the interarm regions that host the magnetic arms of the ordered magnetic fields. 
The selected regions are presented in Table~\ref{names} and Fig.~\ref{m83xregs}.
The left panel of Fig.~\ref{m83recon} shows an overlay of contours of the total power radio emission at 4.86\,GHz on the `total power' X-ray emission in the 
energy band 0.2-2\,keV, smoothed with a Gaussian profile to the same resolution of 15\arcsec. The green lines mark polarisation B-vectors 
and their length is proportional to the degree of polarisation. For the contours of the total power radio map the noise level of the radio polarised intensity 
map was used to better represent weaker levels of the latter map. It is important to note that for the total power radio map a direct correspondence between the total intensity and the strength of the total 
magnetic field exists. However, the intensity of the background X-ray map is a function of the amount of the emitting hot gas, as well as its temperature. 
It is clearly seen that although the total power radio emission quite closely follows the strongest X-ray emission, the magnetic arms are placed just outside of it.

For all spectral regions we followed our analysis performed for NGC\,6946 \citep{wezgowiec16}. 
Each region was fitted with a complex model that consists of two thermal plasma models to account for the disc and the halo emission, 
respectively and a power-law model to describe the emission from the unresolved point sources \citep[see][for details]{wezgowiec16}. To model the emission from the thermal plasma, we use as before the 
{\sc mekal} model \citep{mewe85,kaastra92}. Due to very high sensitivity of the obtained spectra, two thermal components were not sufficient to obtain precise model fits. We found that an additional third 
thermal plasma component has to be added to the model, as also suggested by the outline of the spectral bins. The physical explanation is that two thermal components are needed for a proper description 
of the emission from the galactic halo. The final models include also an absorption component ({\sc wabs} model) to account for the foreground absorption in the Milky Way. For the model 
of the emission from the galactic core an additional component to account for the internal absorption of the central region was needed. 
All models used in our analysis along with the fitted temperatures of the mekal components and photon indices of the power-law components are presented in Table~\ref{m83xtabr}. 
Here, kT$_1$ and kT$_2$ describe the halo emission while kT$_3$ describes the emission from the hot disc. 

\begin{figure*}[ht]
\begin{center}
\resizebox{0.45\hsize}{!}{\includegraphics[angle=-90]{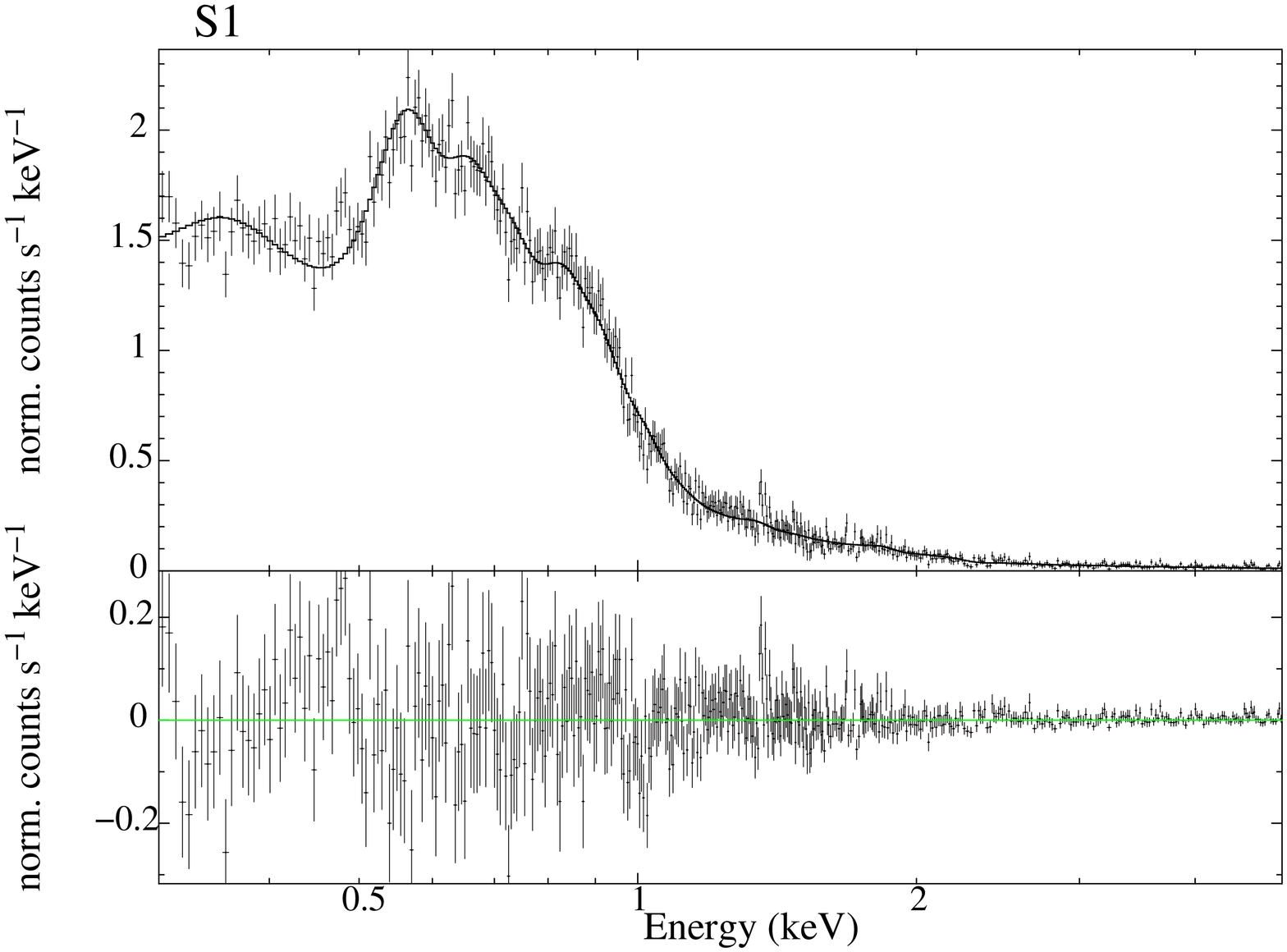}}
\resizebox{0.45\hsize}{!}{\includegraphics[angle=-90]{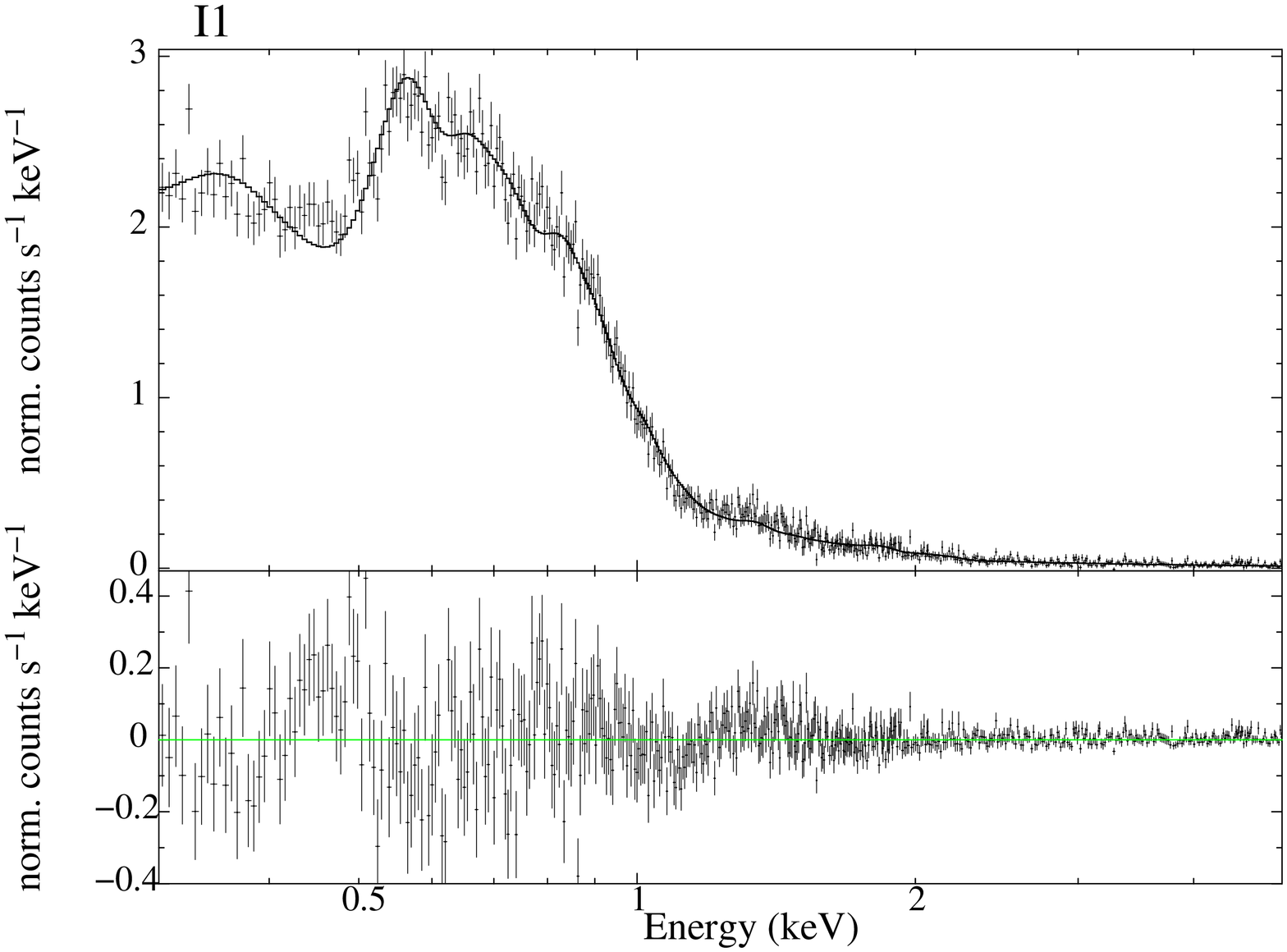}}
\resizebox{0.45\hsize}{!}{\includegraphics[angle=-90]{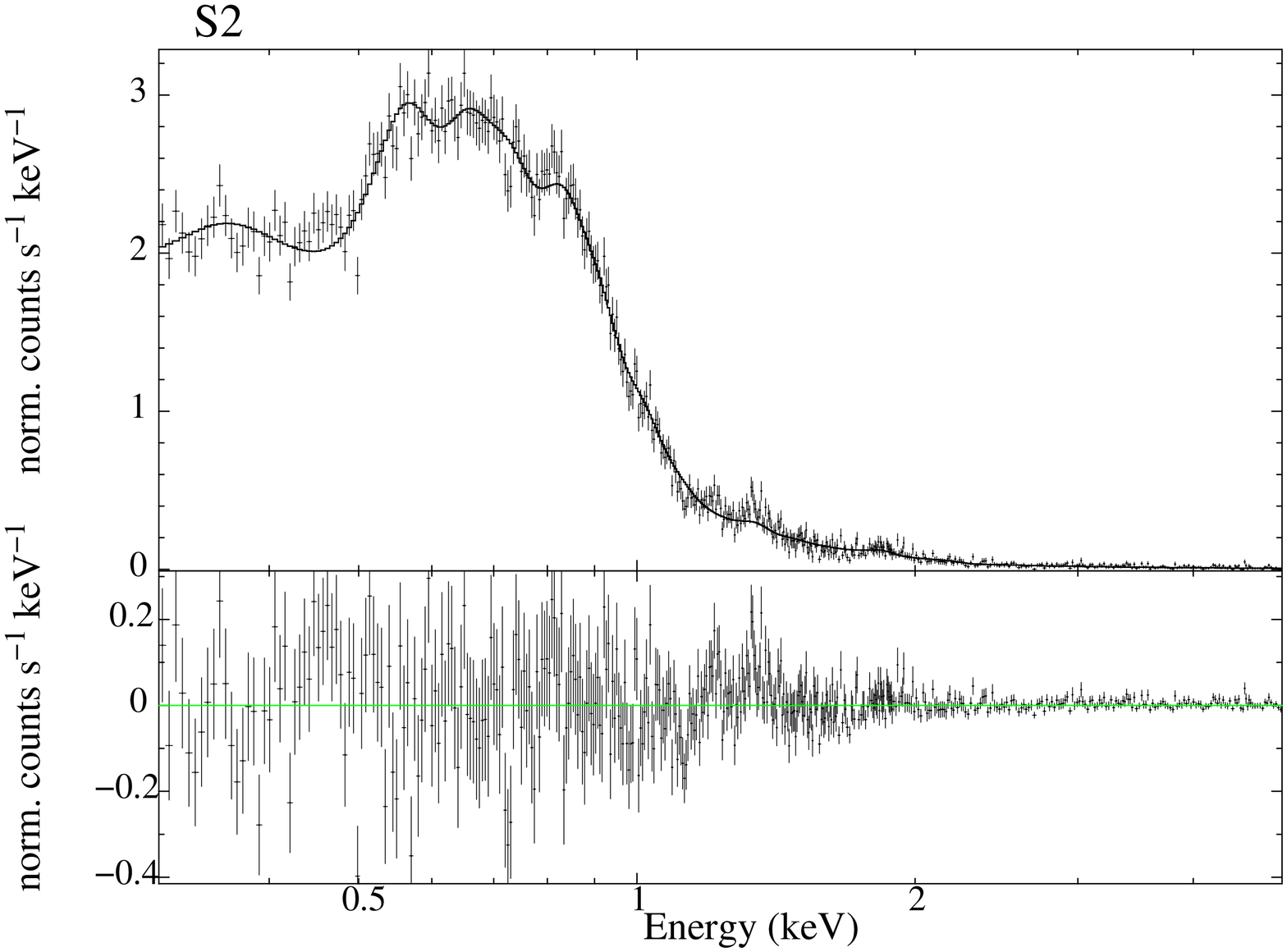}}
\resizebox{0.45\hsize}{!}{\includegraphics[angle=-90]{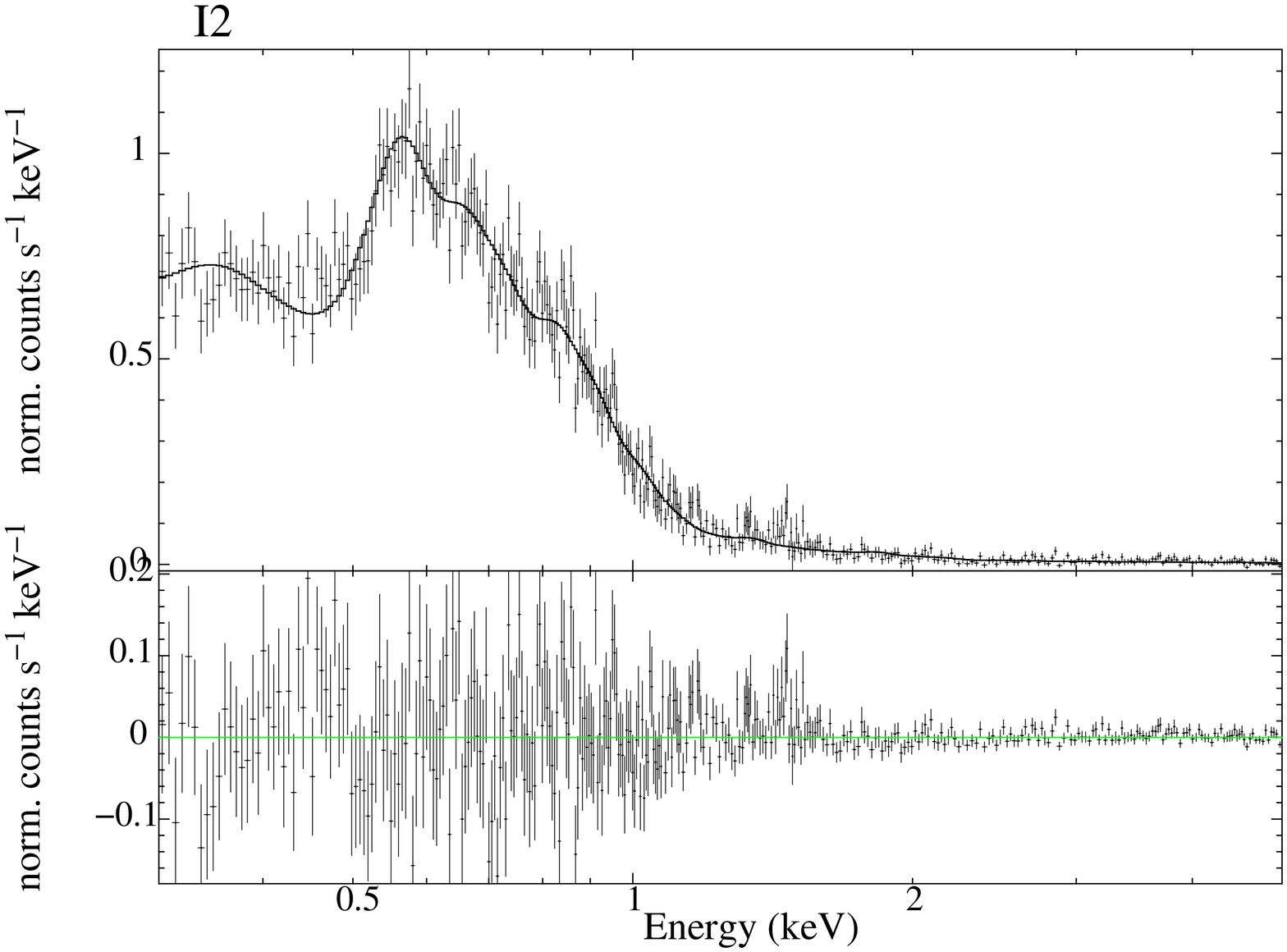}}
\resizebox{0.45\hsize}{!}{\includegraphics[angle=-90]{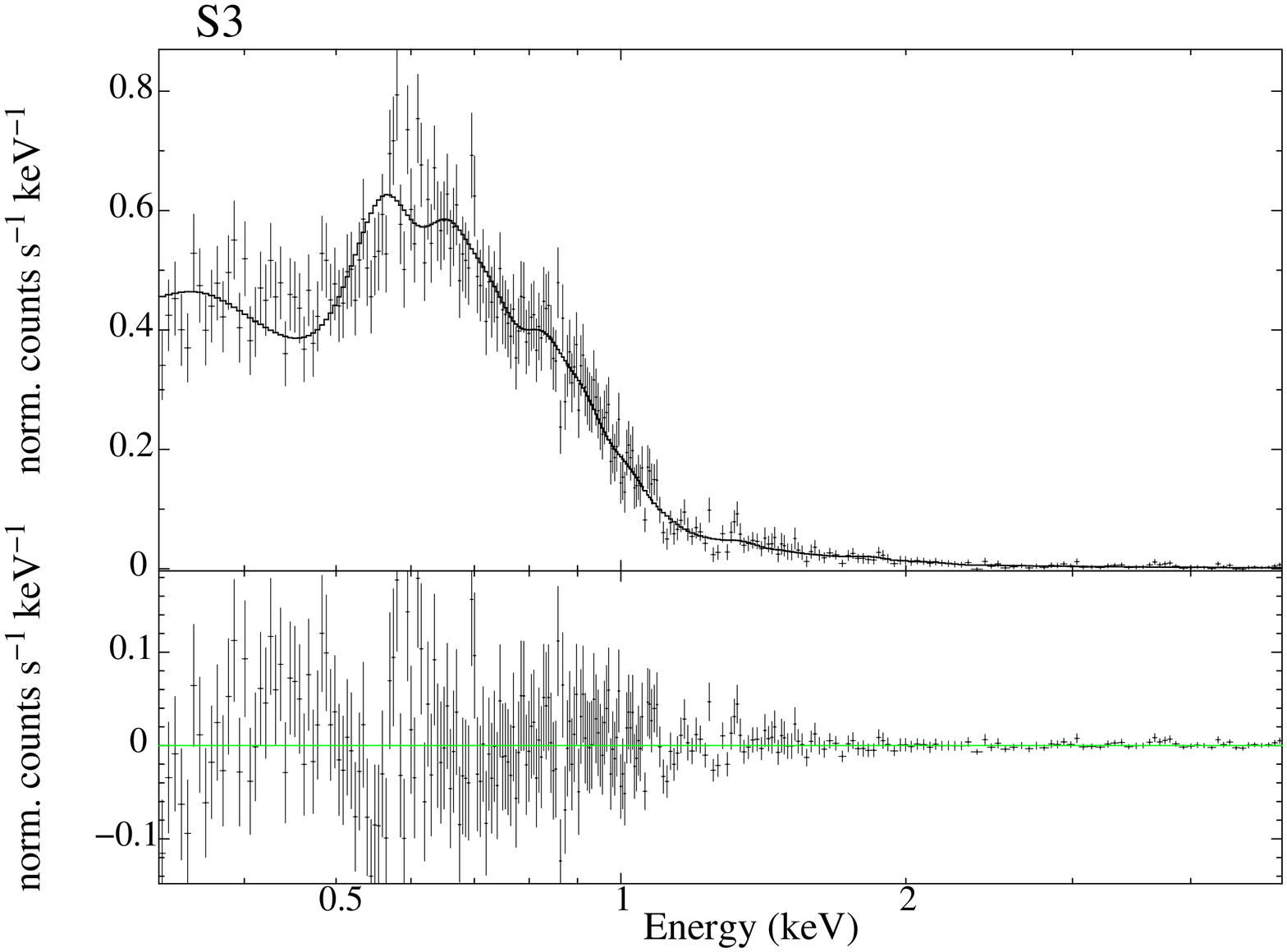}}
\resizebox{0.45\hsize}{!}{\includegraphics[angle=-90]{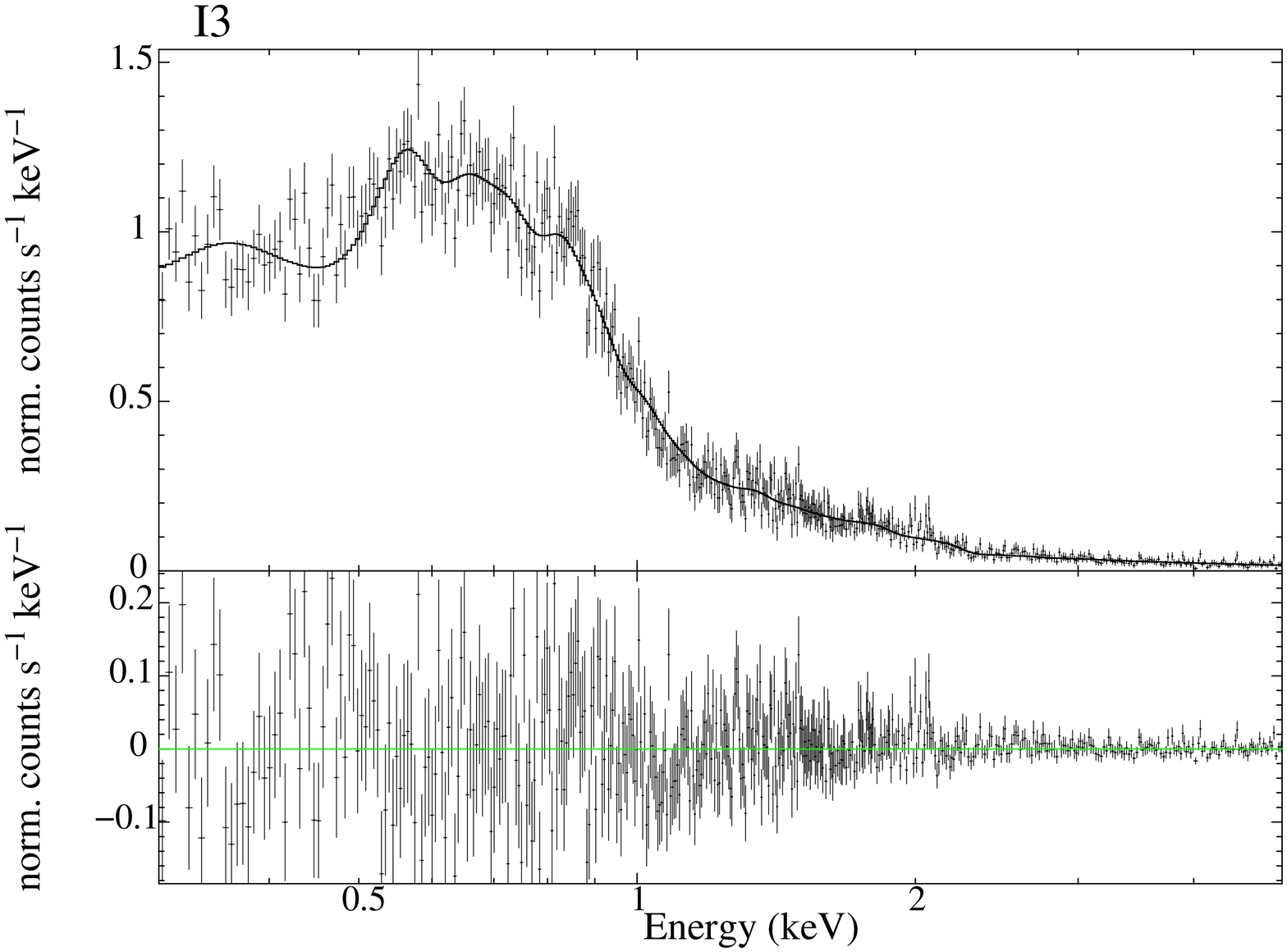}}
\end{center}
\caption{Spectral model fits to the diffuse X-ray emission from the spiral arm (left) and interarm regions (right). See Tables~\ref{m83xtabr} and \ref{m83xfr}.}
\label{m83models}
\end{figure*}

\begin{figure}[ht]
\begin{center}
\resizebox{\hsize}{!}{\includegraphics[angle=-90]{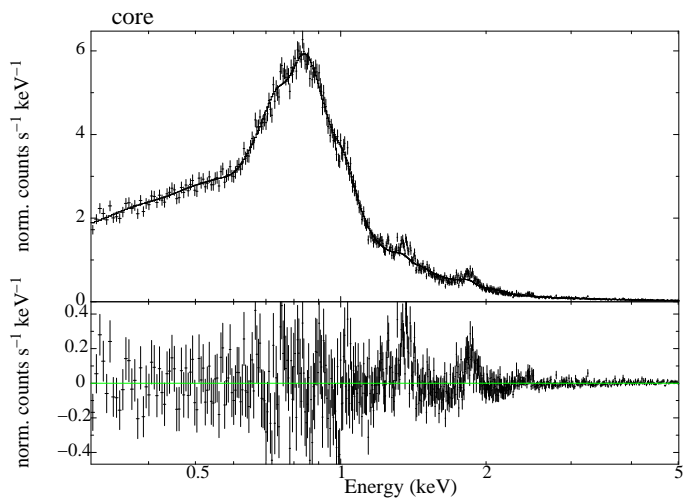}}
\end{center}
\caption{Model fit to the central region of M\,83. See Tables~\ref{m83xtabr} and \ref{m83xfr}.}
\label{m83core}
\end{figure}

The spectra for each spiral and magnetic arm region with the model fit and the residuals are presented in Fig.~\ref{m83models}. The spectrum of the core region is presented in Fig.~\ref{m83core}.
The fluxes derived from the model fits for all regions and spectral components are presented in Table~\ref{m83xfr}.

\begin{table}[ht]
\caption{\label{names}Regions in M\,83 used for the spectral analysis.}
\centering
\begin{tabular}{cl}
\hline\hline
Region name 	& Region description                       \\
\hline
\vspace{5pt}
S1               & south-eastern arm  \\
\vspace{5pt}
S2               & northern arm		  \\
\vspace{5pt}
S3               & western arm  \\
\vspace{5pt}
I1               & eastern interarm  \\
\vspace{5pt}
I2               & north-western interarm  \\
\vspace{5pt}
I3               & south-western interarm  \\
\vspace{5pt}
core             & central region  \\
\hline
\end{tabular}
\end{table}

\begin{table*}[ht]
\caption{\label{m83xtabr}Model fit parameters for the regions studied in M\,83.}
\centering
\begin{tabular}{clccccrr}
\hline\hline
Region      & Model                                 		     & kT$_1$                & kT$_2$                 & kT$_3$      	       & Photon                 &$\chi_{\rm red}^2$\\
	    & type                                  		     & [keV]                 & [keV]                  & [keV]      	       & Index                  &                  \\
\hline
\vspace{5pt}
S1          & wabs(mekal+mekal+mekal+power law)     & 0.10$\pm$0.01 	     & 0.27$^{+0.01}_{-0.02}$ & 0.71$^{+0.01}_{-0.05}$ & 1.77$\pm$0.10		&       1.03       \\
\vspace{5pt}
S2          & wabs(mekal+mekal+mekal+power law)     & 0.11$\pm$0.01	     & 0.28$^{+0.01}_{-0.02}$ & 0.64$^{+0.04}_{-0.03}$ & 2.27$^{+0.11}_{-0.12}$	&       1.29       \\
\vspace{5pt}
S3	    & wabs(mekal+mekal+mekal+power law)     & 0.10$\pm$0.01	     & 0.27$\pm$0.02	      & 0.73$^{+0.11}_{-0.07}$ & 1.75$^{+0.44}_{-0.41}$ &	0.98	   \\
\vspace{5pt}
I1          & wabs(mekal+mekal+mekal+power law)     & 0.10$\pm$0.1	     & 0.28$^{+0.02}_{-0.01}$ & 0.68$\pm$0.04	       & 1.67$\pm$0.10		&       1.17       \\
\vspace{5pt}
I2          & wabs(mekal+mekal+mekal+power law)     & 0.11$^{+0.01}_{-0.02}$ & 0.27$^{+0.02}_{-0.03}$ & 0.69$^{+0.06}_{-0.09}$ & 1.32$^{+0.30}_{-0.28}$	&       1.06       \\
\vspace{5pt}
I3          & wabs(mekal+mekal+mekal+power law)     & 0.11$^{+0.01}_{-0.02}$ & 0.28$^{+0.02}_{-0.03}$ & 0.62$^{+0.07}_{-0.04}$ & 1.63$\pm$0.06		&       1.05       \\
\vspace{5pt}
core        & wabs(mekal+mekal+mekal+wabs\tablefootmark{a}*power law)& 0.12$^{+0.02}_{-0.01}$ & 0.56$^{+0.02}_{-0.01}$ & 0.96$^{+0.06}_{-0.05}$ & 2.27$^{+0.06}_{-0.05}$	&       1.55       \\
\hline
\end{tabular}
\tablefoot{
\tablefoottext{a}{Internal column density of 7.91$^{+1.89}_{-1.63}$$\times$10$^{20}$\,cm$^{-2}$.}
}
\end{table*}

\begin{table*}[ht]
\caption{\label{m83xfr}Total (0.3 - 12 keV) unabsorbed fluxes in 10$^{-14}$erg\,cm$^{-2}$s$^{-1}$ for modelled regions in M\,83.}
\centering
\begin{tabular}{cccccc}
\hline\hline
Region       & mekal 1                       & mekal 2                        & mekal 3	     	               & power law                      & total                      \\
\hline
S1           & 9.65$^{+4.81}_{-3.65}$ (0.19) & 12.96$^{+1.24}_{-1.56}$ (0.26) & 4.44$^{+1.27}_{-0.51}$ (0.09)  & 22.65$^{+4.02}_{-3.41}$ (0.46) & 49.69$^{+11.34}_{-9.13}$   \\
\vspace{5pt}
S2           & 13.04$^{+3.59}_{-5.26}$ (0.18)& 22.47$^{+2.55}_{-2.36}$ (0.31) & 12.27$^{+2.19}_{-2.06}$ (0.17) & 25.36$^{+3.38}_{-2.79}$ (0.34) & 73.18$^{+11.72}_{-12.48}$  \\
\vspace{5pt}
S3	     & 3.63$^{+4.52}_{-1.74}$ (0.23) & 5.81$^{+0.98}_{-0.85}$ (0.38)  & 1.41$^{+0.51}_{-0.28}$ (0.09)  & 4.61$^{+5.19}_{-2.18}$ (0.30)	& 15.46$^{+11.20}_{-5.04}$   \\
\vspace{5pt}
I1           & 19.79$^{+5.55}_{-4.28}$ (0.24)& 21.68$^{+1.92}_{-1.80}$ (0.27) & 7.54$^{+1.39}_{-1.78}$ (0.09)  & 32.05$^{+6.71}_{-5.37}$ (0.40) & 81.07$^{+15.57}_{-13.23}$  \\
\vspace{5pt}
I2           & 6.21$^{+3.48}_{-2.71}$ (0.25) & 8.03$^{+1.22}_{-1.42}$ (0.34)  & 2.01$^{+1.29}_{-0.57}$ (0.08)  & 8.13$^{+7.45}_{-3.76}$ (0.33)  & 24.39$^{+13.44}_{-8.46}$   \\
\vspace{5pt}
I3           & 5.32$^{+2.94}_{-2.26}$ (0.10) & 6.96$^{+1.13}_{-1.25}$ (0.14)  & 3.48$^{+1.01}_{-1.05}$ (0.07)  & 35.59$^{+4.63}_{-3.99}$ (0.69) & 51.35$^{+9.71}_{-8.55}$    \\
\vspace{5pt}
core         & 5.62$^{+5.15}_{-2.83}$ (0.04) & 35.29$^{+1.95}_{-1.21}$ (0.25) & 11.93$^{+1.60}_{-2.02}$ (0.08) & 89.04$^{+7.39}_{-6.41}$ (0.63) & 141.88$^{+16.08}_{-12.47}$ \\
\hline
\end{tabular}
\tablefoot{
Values in brackets are fractions of the total flux for a given component.
}
\end{table*}

\section{Discussion}
\label{disc}

\subsection{Halo and disc components}
\label{components}

The spectral analysis of the available XMM-Newton X-ray data for M\,83 yielded a very interesting result. As mentioned in Sect.~\ref{spectra}, an already complex model of two thermal components 
(with an additional power-law component) did not allow to fit the data properly. Of course, when studying the emission that origins in both the disc and the halo (hence from a face-on galaxy), 
mixing of the gas should lead to a continuous temperature gradient. Due to limited resolution and sensitivity of the studied spectra, a simplified approach needs to be used. Consequently,
distinct temperature components are added to the model, based on the characteristics of the spectra. While typically two thermal components are used in the model, one for the disc and one for the halo, 
in the case of the sensitive observations of M\,83, a third thermal component had to be added to the model. After this modification, the spectra were fitted 
with reduced $\chi^2$ of the order of unity and with errors of the fitted parameters often below ten per cent. The physical interpretation of such model is that the highest temperature 
is associated with the emission from the hot disc and the two lower temperatures are attributed to the X-ray halo. The temperatures of around 0.1\,keV and 0.3\,keV for these two components correspond very well 
to previous results of studies of edge-on spiral galaxies \citep[e.g.][]{strickland04,tuellmann06}, where two thermal components were used to describe the layers of the halo seen from the side.
The hotter of the two halo components can be regarded as a layer of the halo, where the disc gas rises and mixes with the cooler halo gas, being the second thermal component of the halo (upper halo layers).
Nevertheless, when studying the halo of a face-on galaxy, it is very difficult to estimate the sizes and volumes which both components occupy, especially that differences between the halo above the spiral 
arms or interarm regions can be expected. Therefore, in our calculations for the gaseous halo we use the combined model fit values for both halo thermal components. This was obtained by averaging the temperatures with the use of normalisations from the model fits (see Table~\ref{halotemps})
and adding the normalisations and the derived fluxes. 

The extent of the gaseous hot halo of M\,83 was assumed to 20\,kpc to keep the consistency with our analysis of the emission from NGC\,6946 \citep{wezgowiec16}. 
Although for the latter galaxy an extent of 10\,kpc was used, being a typical scale of hot halos visible in edge-on galaxies \citep[e.g.][]{tuellmann06}, 
the Faraday depolarisation map of M\,83 (right panel of Fig.~\ref{m83rm}) suggests that the halo exent could be even twice as large, as the Faraday depolarisation is stronger by a factor of two throughout the central parts 
of the galaxy, when compared to NGC\,6946. Similar values for M\,83 were obtained by \citet{neininger93}, who studied depolarisation in detail using the sensitive low-resolution Effelsberg observations. 
This confirms that the strong Faraday depolarisation occurs globally in the central disc of M\,83. 
If we assume similar gas number densities and strengths of the magnetic field for M\,83 and NGC\,6946, both being starburst galaxies,
this depolarisation needs to be caused by a stronger Faraday Rotation (by a factor of two) resulting from the twice longer pathlength through the halo of M\,83.
This is what we observe in the Rotation Measure maps \citep[left panel of Fig.~\ref{m83rm} and the left panel of Fig.~12 in][]{beck07}. 
Of course, most of the depolarisation occurs in the halo gas that is much denser than the hot X-ray gas. However, with an assumption of a simple scaling of the halo extent, 
a depolarisation layer that is two times thicker would infer a similar change in the extent of the hot gas component. Our assumption of the extent of the halo of M\,83 can 
be also verified with the star-forming properties of both galaxies. A significant star formation in the underlying disc certainly is an important driver of an extended gaseous halo. Also in this aspect M\,83 and NGC\,6946 differ significantly. 
While both galaxies show extended discs in the far-infrared maps, the total flux density of M\,83 is around two times higher than that of NGC\,6946 \citep{fitt92}, which, taking into account the distances 
and angular sizes of both galaxies, translates into the surface star-formation rates over 1.6 times higher in M\,83 than in NGC\,6946.
Also the recent studies of star-forming properties of a sample of nearby galaxies by \citet{tan18} confirm significant differences between M\,83 and NGC\,6946, especially that both galaxies are in their work 
assumed to be at almost the same distances and of similar sizes. 

\begin{table}[ht]
\caption{\label{halotemps}Temperatures of the hot gas in the halo and in the disc of M\,83.}
\centering
\begin{tabular}{cll}
\hline\hline
Region      & kT$_{halo}$\tablefootmark{a} [keV] & kT$_{disc}$\tablefootmark{b} [keV] \\
\hline
\vspace{5pt}
S1          & 0.14$\pm$0.01 		& 0.71$^{+0.01}_{-0.05}$ \\
\vspace{5pt}
S2          & 0.17$\pm$0.01 		& 0.64$^{+0.04}_{-0.03}$ \\
\vspace{5pt}
S3          & 0.15$\pm$0.01 		& 0.73$^{+0.11}_{-0.07}$ \\
\vspace{5pt}
I1          & 0.14$\pm$0.01 		& 0.68$\pm$0.04 \\
\vspace{5pt}
I2          & 0.16$^{+0.01}_{-0.02}$ 	& 0.69$^{+0.06}_{-0.09}$ \\
\vspace{5pt}
I3          & 0.16$^{+0.01}_{-0.02}$ 	& 0.62$^{+0.07}_{-0.04}$ \\
\vspace{5pt}
core        & 0.41$^{+0.04}_{-0.01}$ 	& 0.96$^{+0.06}_{-0.05}$ \\
\hline
\end{tabular}
\tablefoot{
\tablefoottext{a}{Normalisation-weighted averages of kT$_1$ and kT$_2$ from Table~\ref{m83xtabr}.}
\tablefoottext{b}{kT$_3$ from Table~\ref{m83xtabr}.}
}
\end{table}

\subsection{Parameters of the hot gas}
\label{gasparams}

In Table~\ref{m83xfr}, where the unabsorbed fluxes of the model fit components for each region are presented, in brackets the contribution of a given component to the total flux is shown.
We focus only on the thermal components of our model fits that correspond to the hot gas in the disc and in the halo of M\,83.
For all regions except S2 roughly the same contribution to the total flux (around eight per cent) comes from the hottest thermal component that is assumed to be 
associated with the disc gas. In the region S2, however, the contribution from the disc gas is higher by a factor of two, which suggests that more efficient mixing of the disc and the halo gas occurs. 
This is confirmed by the second lowest temperature of the disc gas and the highest temperature of the halo gas among all spectral regions except the core, where overally higher temperatures
are expected (Table~\ref{halotemps}). The lowest average temperatures of the hot gas (in the disc and in the halo) in the region S2 finds its reflection also in the hardness ratio map (Fig.~\ref{m83hr}).

We used the model fit parameters (see Sect.~\ref{spectra}) to calculate number densities, gas masses, thermal energies and their densities, as well as the cooling times of the hot gas for all regions 
that we used in our spectral analysis. Our calculations are based on the model of thermal cooling and ionisation equilibrium of \citet{nulsen84}, where $L_X=1.11\cdot \Lambda(T)\,n^2_e\,V\,\eta$,
$\eta$ is an unknown filling factor and $\Lambda(T)$ is a cooling coefficient of the order of $10^{-22}\,{\rm erg}\,{\rm cm}^3\,s^{-1}$ for temperatures of a few millions~K
\citep{raymond76}. In Sect.~\ref{components} we described the need of averaging the parameters for the model components associated with the hot gas in the galactic halo. 
As a result, in this section we discuss the properties of the hot gas in the disc (Table~\ref{disk}) and in the halo (Table~\ref{halo}) of M\,83. We do not focus on the central (core) region, because 
the values of the hot gas parameters are, as expected, always extreme and should not be considered when a comparison between the spiral and the interarm regions is presented. These values, however, 
are presented for the completeness of the global picture of the hot gas in M\,83.

The parameters of the hot gas are relatively constant throughout the disc of M\,83. The only exception is the spiral arm S2. In this region the number density, derived directly from 
the spectral fit and therefore independent of the emitting volume, is significantly higher. 
It could be likely explained that this region corresponds to the most prominent star-forming arm in the galaxy.

The H$\alpha$ image of M\,83 clearly shows that the interarm region I1 is still significantly `contaminated' with the star-forming regions. 
Consequently, it may resemble a weak spiral arm (like the spiral arm S3) 
rather than a distinct interarm region. The interarm regions I2 and I3 are relatively free from the star-forming regions and thus a comparison with the spiral arms (especially S1 and S2) is more reliable. The number densities of the hot gas and the densities of the thermal energies are visibly lower in the interarm regions (Table~\ref{disk}).

The properties of the hot gas in the halo of M\,83 reflect the findings for the disc gas, but 
the highest number density above the interarm region I1 is quite surprising (Table~\ref{halo}).

\begin{table*}[ht]
\caption{\label{disk}Derived parameters of the hot gas in the disc of M\,83.}
\centering
\begin{tabular}{cccccc}
\hline\hline
Region      & n$^{disc}\eta^{-0.5}$     & M$^{disc}_{gas}\eta^{0.5}$       & E$^{disc}_{th}\eta^{0.5}$   & $\epsilon^{disc}_{th}\eta^{-0.5}$    & ($\tau^{disc}\eta^{0.5}$    \\
\vspace{5pt}& [10$^{-3}$cm$^{-3}$]   & [10$^6$M$_\odot$]             & [10$^{54}$\,erg]         & [10$^{-12}$\,erg\,cm$^{-3}$]      & [Myr]                    \\
\hline
\vspace{5pt}
S1          & 3.66$^{+0.46}_{-0.21}$ & 1.00$^{+0.12}_{-0.55}$        & 2.01$^{+0.28}_{-0.25}$   & 6.26$^{+0.89}_{-0.77}$            & 552$^{+3}_{-61}$         \\
\vspace{5pt}
S2          & 5.69$^{+0.45}_{-0.48}$ & 1.72$^{+0.14}_{-0.15}$        & 3.16$^{+0.46}_{-0.40}$   & 8.77$^{+1.28}_{-1.12}$            & 314$^{+9}_{-15}$         \\
\vspace{5pt}
S3          & 3.13$^{+0.47}_{-0.23}$ & 0.37$^{+0.06}_{-0.03}$        & 0.78$^{+0.25}_{-0.13}$   & 5.51$^{+1.79}_{-0.89}$            & 671$^{+18}_{-31}$        \\
\vspace{5pt}
I1          & 3.58$^{+0.29}_{-0.43}$ & 1.69$^{+0.14}_{-0.20}$        & 3.30$^{+0.48}_{-0.56}$   & 5.86$^{+0.85}_{-1.00}$            & 533$^{+18}_{-47}$        \\
\vspace{5pt}
I2          & 2.46$^{+0.65}_{-0.35}$ & 0.66$^{+0.18}_{-0.09}$        & 1.31$^{+0.49}_{-0.33}$   & 4.09$^{+1.54}_{-1.03}$            & 792$^{+128}_{-34}$       \\
\vspace{5pt}
I3          & 3.42$^{+0.44}_{-0.53}$ & 0.80$^{+0.10}_{-0.13}$        & 1.42$^{+0.36}_{-0.30}$   & 5.10$^{+1.30}_{-1.07}$            & 499$^{+13}_{-65}$        \\
\vspace{5pt}
core        & 18.0$^{+0.8}_{-1.2}$   & 0.64$^{+0.03}_{-0.04}$        & 1.76$^{+0.19}_{-0.21}$   & 41.6$^{+4.6}_{-4.8}$              & 180$^{+4}_{-12}$         \\
\hline
\end{tabular}
\tablefoot{
The columns are the region name, electron number density, total gas mass, total thermal energy, thermal energy density, and cooling time. $\eta$ is the volume filling factor.
}
\end{table*}

\begin{table*}[ht]
\caption{\label{halo}Derived parameters of the hot gas in the halo of M\,83.}
\centering
\begin{tabular}{cccccc}
\hline\hline
Region 	    & n$^{halo}\eta^{-0.5}$     & M$^{halo}_{gas}\eta^{0.5}$       & E$^{halo}_{th}\eta^{0.5}$   & $\epsilon^{halo}_{th}\eta^{-0.5}$    & ($\tau^{halo}\eta^{0.5}$    \\
\vspace{5pt}& [10$^{-3}$cm$^{-3}$]   & [10$^6$M$_\odot$]             & [10$^{54}$\,erg]         & [10$^{-12}$\,erg\,cm$^{-3}$]      & [Myr]          	       \\
\hline
\vspace{5pt}
S1          & 3.10$^{+0.28}_{-0.31}$ & 16.7$^{+1.5}_{-1.7}$          & 6.7$^{+1.1}_{-0.7}$  	& 1.04$^{+0.18}_{-0.11}$  	    & 361$^{+28}_{-61}$        \\
\vspace{5pt}
S2          & 3.26$^{+0.22}_{-0.24}$ & 19.7$^{+1.3}_{-1.5}$          & 9.6$\pm$1.2	  	& 1.33$\pm$0.17		  	    & 330$^{+13}_{-36}$       \\
\vspace{5pt}
S3          & 3.07$^{+0.65}_{-0.52}$ & 7.3$^{+1.5}_{-1.2}$	     & 3.1$\pm$0.7	  	& 1.11$^{+0.24}_{-0.25}$  	    & 403$^{+94}_{-25}$      \\
\vspace{5pt}
I1          & 3.38$^{+0.19}_{-0.21}$ & 31.8$^{+1.8}_{-1.9}$          & 12.8$^{+1.7}_{-0.8}$  	& 1.13$^{+0.15}_{-0.07}$  	    & 375$^{+15}_{-38}$        \\
\vspace{5pt}
I2          & 2.32$^{+0.33}_{-0.16}$ & 12.5$^{+1.8}_{-0.8}$          & 5.7$^{+0.8}_{-1.1}$  	& 0.89$^{+0.13}_{-0.16}$  	    & 489$^{+69}_{-74}$       \\
\vspace{5pt}
I3          & 2.32$^{+0.33}_{-0.18}$ & 10.9$^{+1.5}_{-0.9}$          & 5.0$^{+0.7}_{-1.0}$  	& 0.89$^{+0.13}_{-0.17}$  	    & 495$^{+70}_{-62}$       \\
\vspace{5pt}
core	    & 7.59$^{+0.71}_{-0.61}$ & 5.4$^{+0.5}_{-0.4}$           & 6.4$^{+1.3}_{-0.7}$      & 7.50$^{+1.51}_{-0.77}$            & 190$^{+1}_{-4}$       \\
\hline
\end{tabular}
\tablefoot{
The columns are the region name, electron number density, total gas mass, total thermal energy, thermal energy density, and cooling time. $\eta$ is the volume filling factor.
}
\end{table*}

\subsection{Magnetic fields in M\,83}
\label{bfields}

We used both total power and polarised intensity radio maps at 4.86\,GHz ($\lambda$6cm) to measure fluxes with the resolution of 15$\arcsec$ in all spiral arm and interarm regions.
For the calculation of the strengths and energy densities of both the total and ordered magnetic fields we used the energy equipartition formula provided by 
\citet{beck05}. Assuming a synchrotron spectral index of 0.8, an inclination of the galactic disc of 15$\degr$ (Table~\ref{astrdat}), and a proton-to-electron ratio of 100, we calculated 
the parameters of the magnetic fields for a galactic disc thickness of 1\,kpc.
Because the main uncertainty is related to both proton-to-electron ratio and the pathlength, we performed the calculations varying these parameters by 50\%. The resulting 
errors are of the order of 30\% but since these are systematic errors, the real uncertainties are, in fact, lower. The results of our calculations are presented in Table~\ref{magparams}.

The strongest magnetic fields are detected in the prominent spiral arms S1 and S2. The magnetic field strength in the weak spiral arm S3 is comparable to that found 
in all interarm regions. Consequently, all these four regions show lower energy densities of the magnetic field. 
The strengths of the ordered magnetic field are significantly higher in the magnetic arms than in the spiral arms. As a result, 
the degrees of polarisation are in these areas higher than in the spiral arms by a factor of two (or over three in the case of the spiral arm S2).

\begin{table*}[ht]
\caption{\label{magparams} Properties of the magnetic fields in M\,83.}
\centering
\begin{tabular}{cccccc}
\hline\hline
Region          &S$_{synch}$    &p$_{synch}$    &B$_{tot}$      	&$\epsilon_B$                   	&B$_{ord}$	   \\
\vspace{5pt}    &[mJy/beam]     &[\%]           &[$\mu$G]       	&[10$^{-12}$\,erg\,cm$^{-3}$]   	&[$\mu$G]	   \\
\hline
\vspace{5pt}
S1              &  0.71         &   11.4        &  16.2$\pm$5.4         &   10.5$\pm$6.0                        &  4.8$\pm$1.6     \\
\vspace{5pt}
S2              &  0.91         &    7.1        &  17.6$\pm$5.9         &   12.3$\pm$7.0                        &  4.0$\pm$1.3     \\
\vspace{5pt}
S3              &  0.43         &   11.7        &  14.2$\pm$4.8         &    8.0$\pm$5.5                        &  4.3$\pm$1.4     \\
\vspace{5pt}
I1              &  0.58         &   22.7        &  14.8$\pm$5.0         &    8.7$\pm$5.2                        &  6.7$\pm$2.2     \\
\vspace{5pt}
I2		&  0.42         &   25.8        &  13.5$\pm$4.5         &    7.2$\pm$4.5                        &  6.6$\pm$1.9     \\
\vspace{5pt}
I3              &  0.51         &   21.8        &  14.4$\pm$4.8         &    8.2$\pm$5.3                        &  6.3$\pm$1.7     \\
\hline
\end{tabular}
\tablefoot{
The columns are the region name, non-thermal radio flux, degree of polarisation, total magnetic field strength, magnetic field energy, and ordered magnetic field strength.
}
\end{table*}

\subsection{Hot gas and magnetic fields}
\label{hotfields}

Since in the case of a face-on galaxy the information about its magnetic fields comes from the observations of the radio emission integrated along the line of sight, we cannot convincingly disentangle 
radio emission from the disc and that from the halo. This can be done for edge-on galaxies and some estimates of the magnetic field strengths in the galactic halo are possible, as shown 
recently by \citet{stein19}. 
Of course, some estimates of the line-of-sight component can be done using 
the information provided by the Faraday Rotation or the depolarisation maps. This information, however, we used only to estimate the vertical extent of the halo of M\,83 (see Sect.~\ref{components}). 
Therefore, to compare the properties of the magnetic fields in M\,83 with 
properties of its hot gas \citep[and follow the analysis presented in][]{wezgowiec16}, we needed to calculate the global properties of the hot gas by combining the information 
obtained for both the disc and the halo.

While for the gas masses and energies the global values are just the sums of the disc and the halo values, global or average number densities were calculated as an average of the disc and the halo number densities with the emitting volumes used as the weights. Of course, because the extent of the halo was assumed as large as 20\,kpc, the final values of 
the number density did not significantly deviate from that of the halo only (see Tables~\ref{halo} and \ref{totalhot}). The final energy densities were calculated by dividing the sums of energies by sums of volumes of the disc and the halo components.

The global properties of the hot gas, in the specified regions of the disc and the halo above them, could be then compared with the magnetic fields. For a direct comparison, that 
also allowed to study the energy budget of the entire system, we used energy densities of both the hot gas and the magnetic fields. However, because the calculated 
value of the energy density of the hot gas depends on an unknown filling factor $\eta$, which is difficult to estimate, similarly to our study of NGC\,6946 \citep{wezgowiec16}, 
we used a different quantity - energy per particle ($E_p$). It is the ratio of the thermal energy density and the number density of the hot gas, which is independent of the filling factor. 
We compared these values with the energy densities of the magnetic field (Table~\ref{particles}), which allowed to search for fingerprints of the possible reconnection heating, 
found in NGC\,6946 \citep{wezgowiec16}. 

\begin{table}[ht]
\caption{\label{totalhot} Global parameters of the hot gas in the spectral regions of M\,83.}
\centering
\begin{tabular}{ccccc}
\hline\hline
Reg.      & n$^{avg}\eta^{-0.5}$   & M$^{tot}_{gas}\eta^{0.5}$     & E$^{tot}_{th}\eta^{0.5}$ & $\epsilon^{tot}_{th}\eta^{-0.5}$   \\
\vspace{5pt}
	  & [10$^{-3}$cm$^{-3}$]   & [10$^6$M$_\odot$]             & [10$^{54}$\,erg]         & [10$^{-12}$		           \\
\vspace{5pt}
	  &			   &				   &			      & erg\,cm$^{-3}$]			   \\
\hline
\vspace{5pt}
S1          & 3.12$^{+0.29}_{-0.31}$ & 17.7$\pm$1.7	             & 8.7$^{+1.4}_{-0.9}$      & 1.29$^{+0.21}_{-0.14}$           \\
\vspace{5pt}
S2          & 3.38$^{+0.23}_{-0.25}$ & 21.5$^{+1.4}_{-1.6}$          & 12.8$^{+1.7}_{-1.6}$     & 1.69$\pm$0.22                    \\
\vspace{5pt}
S3          & 3.08$^{+0.65}_{-0.50}$ & 7.6$^{+1.6}_{-1.2}$           & 3.9$^{+0.9}_{-0.8}$      & 1.32$^{+0.31}_{-0.28}$           \\
\vspace{5pt}
I1          & 3.39$^{+0.20}_{-0.22}$ & 33.5$^{+1.9}_{-2.1}$          & 16.1$^{+2.2}_{-1.3}$     & 1.36$^{+0.18}_{-0.11}$           \\
\vspace{5pt}
I2          & 2.33$^{+0.34}_{-0.16}$ & 13.1$^{+1.9}_{-0.9}$          & 7.0$^{+1.3}_{-1.4}$      & 1.05$^{+0.19}_{-0.21}$           \\
\vspace{5pt}
I3          & 2.37$^{+0.33}_{-0.20}$ & 11.7$^{+1.7}_{-1.0}$          & 6.4$^{+1.1}_{-1.3}$      & 1.09$^{+0.18}_{-0.22}$           \\
\vspace{5pt}
core        & 8.09$^{+0.72}_{-0.64}$ & 6.1$\pm$0.5	             & 8.1$^{+1.5}_{-0.9}$      & 9.12$^{+1.65}_{-0.96}$           \\
\hline
\end{tabular}
\tablefoot{
The columns are the region name, electron number density, total gas mass, total thermal energy, and thermal energy density. $\eta$ is the volume filling factor.
}
\end{table}

\begin{table}[ht]
\caption{\label{particles}Thermal energy per particle and energy densities of the magnetic field for regions of the spiral and magnetic arms of M\,83.}
\centering
\begin{tabular}{ccc}
\hline\hline
Region      & $E_p$                 & $\epsilon_{B}$\\
\vspace{5pt}
	    & [10$^{-9}$\,erg]	    & [10$^{-12}$\,erg\,cm$^{-3}$]\\
\hline
\vspace{5pt}
S1          & 0.41$^{+0.03}_{-0.04}$& 10.5$\pm$6.0  \\
\vspace{5pt}
S2          & 0.50$\pm$0.03& 12.3$\pm$7.0  \\
\vspace{5pt}
S3          & 0.43$^{+0.01}_{-0.03}$& 8.0$\pm$5.5   \\
\vspace{5pt}
I1          & 0.40$^{+0.03}_{-0.01}$  & 8.7$\pm$5.2 \\
\vspace{5pt}
I2          & 0.45$^{+0.02}_{-0.06}$  & 7.2$\pm$4.5 \\
\vspace{5pt}
I3          & 0.46$^{+0.01}_{-0.06}$  & 8.2$\pm$5.3 \\
\hline
\end{tabular}
\end{table}

The left panel of Fig.~\ref{epmag} presents the energy per particle of the hot gas and the energy density of the magnetic fields in the studied spectral regions of M\,83.
Because the star-forming efficiency varies strongly across the disc, one should expect the highest energies of the hot gas and of the magnetic fields, as a result of intense star formation. The interarm regions, on the other hand, are areas free of any significant star formation and should show lower energies of the hot gas and of the magnetic fields.
The left panel of Fig.~\ref{epmag} and the corresponding Table~\ref{particles} show however, that lower thermal energies per particle and higher energy densities of the magnetic fields 
are observed for the spiral arms. The exceptions are the spiral arm S2 and the interarm region I1.

As already mentioned in Sect.~\ref{gasparams}, the interarm region I1 still includes some of the distinct star-forming regions and therefore can be regarded as a weak spiral (star-forming) 
arm. In this sense, the lowest value of $E_p$ should not be used as representative for an interarm region. In the region S2 the hottest component of the gas is not 
confined strictly to the disc and mixes more efficiently with the halo gas (as already mentioned in Sect.~\ref{gasparams}), 
which causes that the global energy density has a much higher contribution from the dense disc gas. 
Because the region S2 corresponds to the most prominent star-forming spiral arm, that origins already within the galactic bar, 
the observed thermal energy density is the highest among all regions (see Table~\ref{totalhot}). With 
much smaller differences in the derived number densities, this leads to the highest value of energy per particle (Table~\ref{particles}) for this region. 

\subsubsection{Reconnection heating}

\begin{figure*}[ht]
\resizebox{0.44\hsize}{!}{\includegraphics[clip]{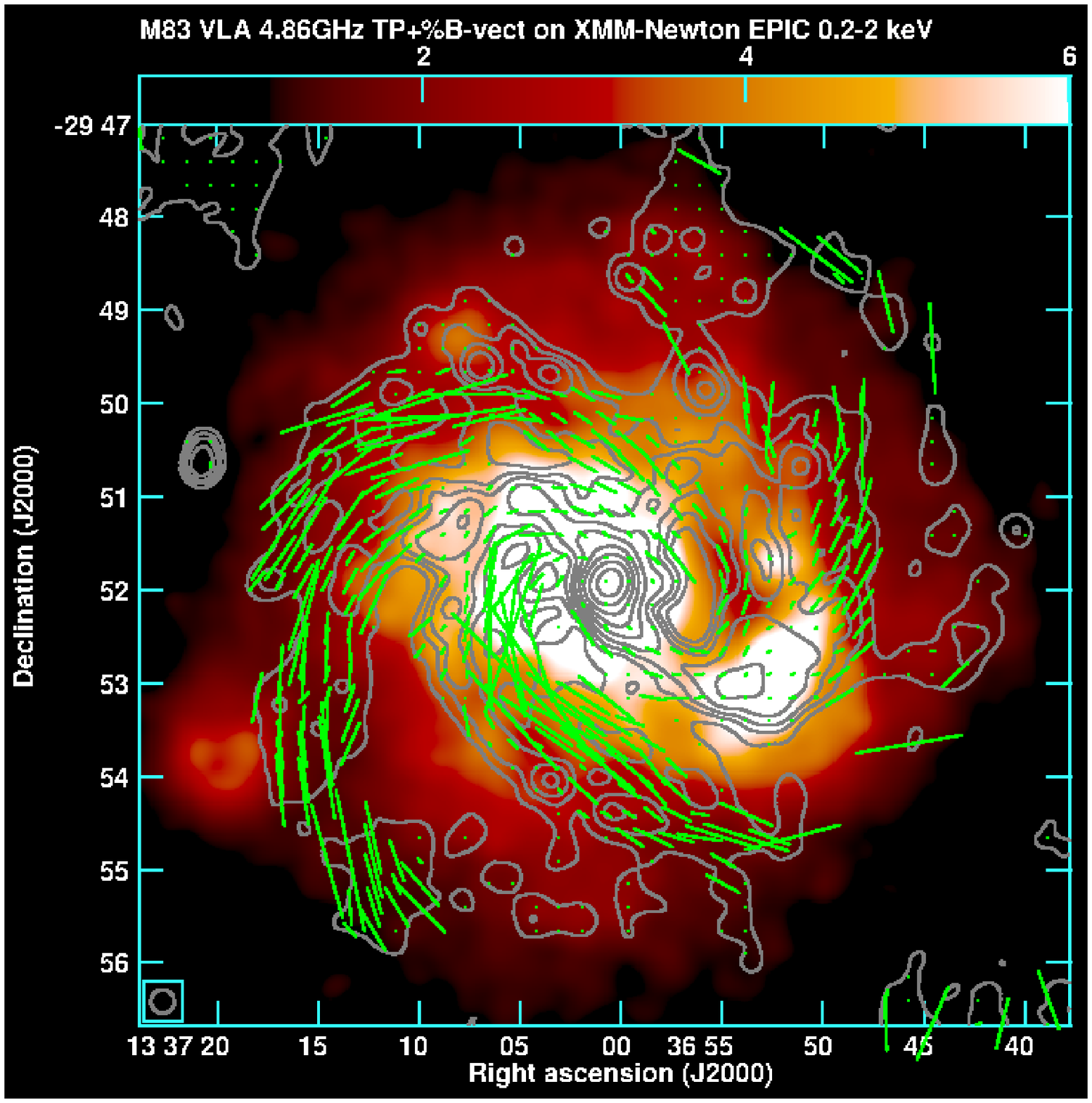}}
\resizebox{0.56\hsize}{!}{\includegraphics[clip]{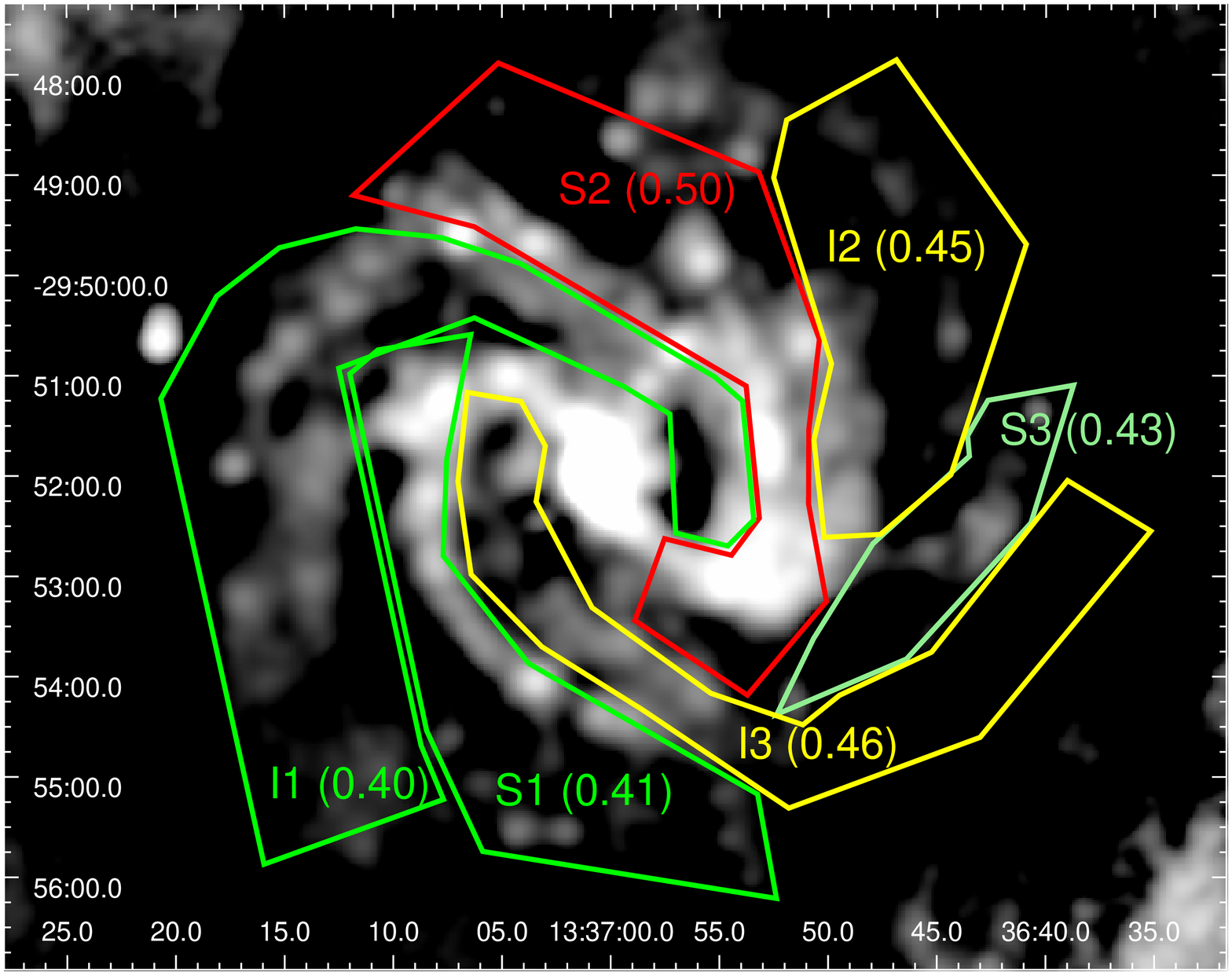}}    
        \caption{
                 Left: Total power radio emission at 4.86\,GHz from M\,83 with vectors of B-field overlaid on the XMM-Newton EPIC X-ray image in the energy band 0.2-2\,keV.
		 Right: Spectral regions with colour-coded (low/green to high/red) energies per particle (in parentheses) overlaid on the total power radio emission at 4.86\,GHz.
        }
\label{m83recon}
\end{figure*}

In the reconnection heating scenario, the energy that comes from the turbulent component of the magnetic field is converted into the thermal energy of the surrounding medium, which 
should be observed as an increase in the temperature of the hot gas. 
Although reconnection should be most efficient in the spiral arms, where the turbulent component of the magnetic field is strong, 
high turbulence and supernova heating most likely lead to removal of any field ordering. Furthermore, 
the slight increase in temperature can easily remain unnoticed due to already efficient heating from the supernovae.
The interarm regions, however, may host conditions that favour the detection of both increasing ordering of the magnetic field and heating of the surrounding ISM.
This seems to be visible in the left panel of Fig.~\ref{m83recon}. The magnetic arms clearly avoid the 
brightest X-ray areas. They are placed directly next to these areas, which would mean that with the decrease in the amount of the hot emitting gas, and thus its density, the 
action of the reconnection effects may become visible as an increase in the regularity of the magnetic field and in heating of the surrounding gas.

If we exclude from the comparison regions S2 and I1 (see above), 
the slight increase in temperature of the hot gas is indeed visible 
in the magnetic arm regions I2 and I3, when compared to the spiral arm regions S1 and S3 (see Table~\ref{onetemps}). A similar trend is visible for energies per particle derived from the model 
fits (see Table~\ref{particles} and the right panel of Fig.~\ref{m83recon}). However, all values for these four regions (I2, I3, S1, and S3) agree within errors. Therefore, this 
result would need further confirmation with future X-ray missions that will offer higher spectral sensitivity.
On the other hand, for each of these regions, in the case of a typical spiral galaxy, we would expect higher temperatures of the hot gas in the spiral arms, that result from the intense local star formation, absent in the interarm regions. Consequently, the highest average temperature of the hot gas was found for the spiral arm S2, 
which shows the most prominent H$\alpha$ emission (Fig.~\ref{m83xregs}). However, the second highest temperature is that for the interarm region I3, that is almost entirely devoid of H$\alpha$ regions. This would mean, that although we did not find any significant temperature increase in the interarm regions of M\,83, 
the comparable values throughout the disc require an additional heating source for the interarm regions.
Here, higher energies per particle with the simultaneous lower energies of the magnetic field (right panel of Fig.~\ref{m83recon} and Table~\ref{particles}) would still agree with the reconnection heating scenario 
of a conversion of the magnetic energy into the thermal energy of the gas. Such conversion would be done via the reconnection processes, which was also postulated for NGC\,6946 \citep{wezgowiec16}.

\begin{figure*}[ht]
\resizebox{0.5\hsize}{!}{\includegraphics[clip,angle=-90]{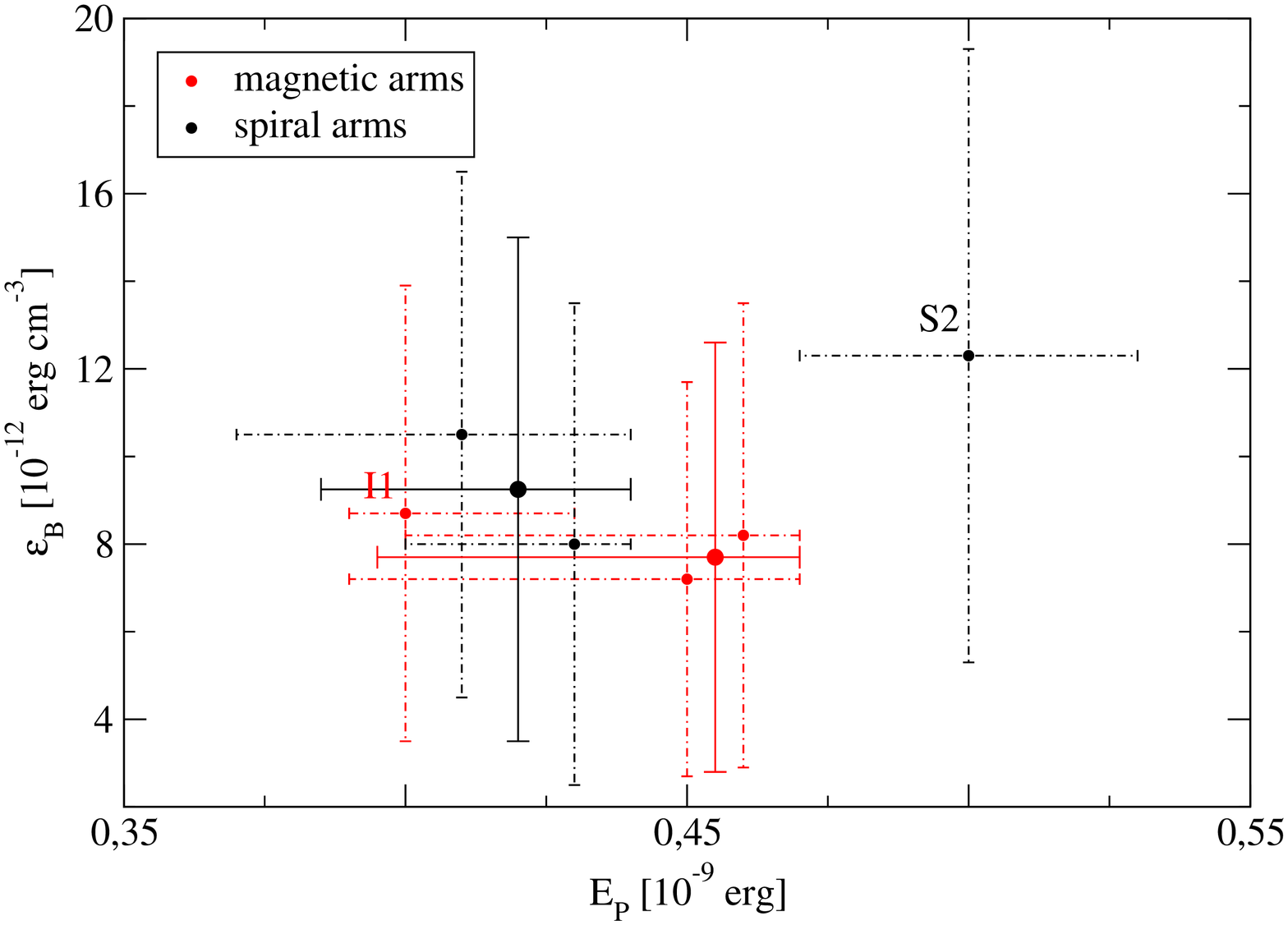}}
\resizebox{0.5\hsize}{!}{\includegraphics[clip,angle=-90]{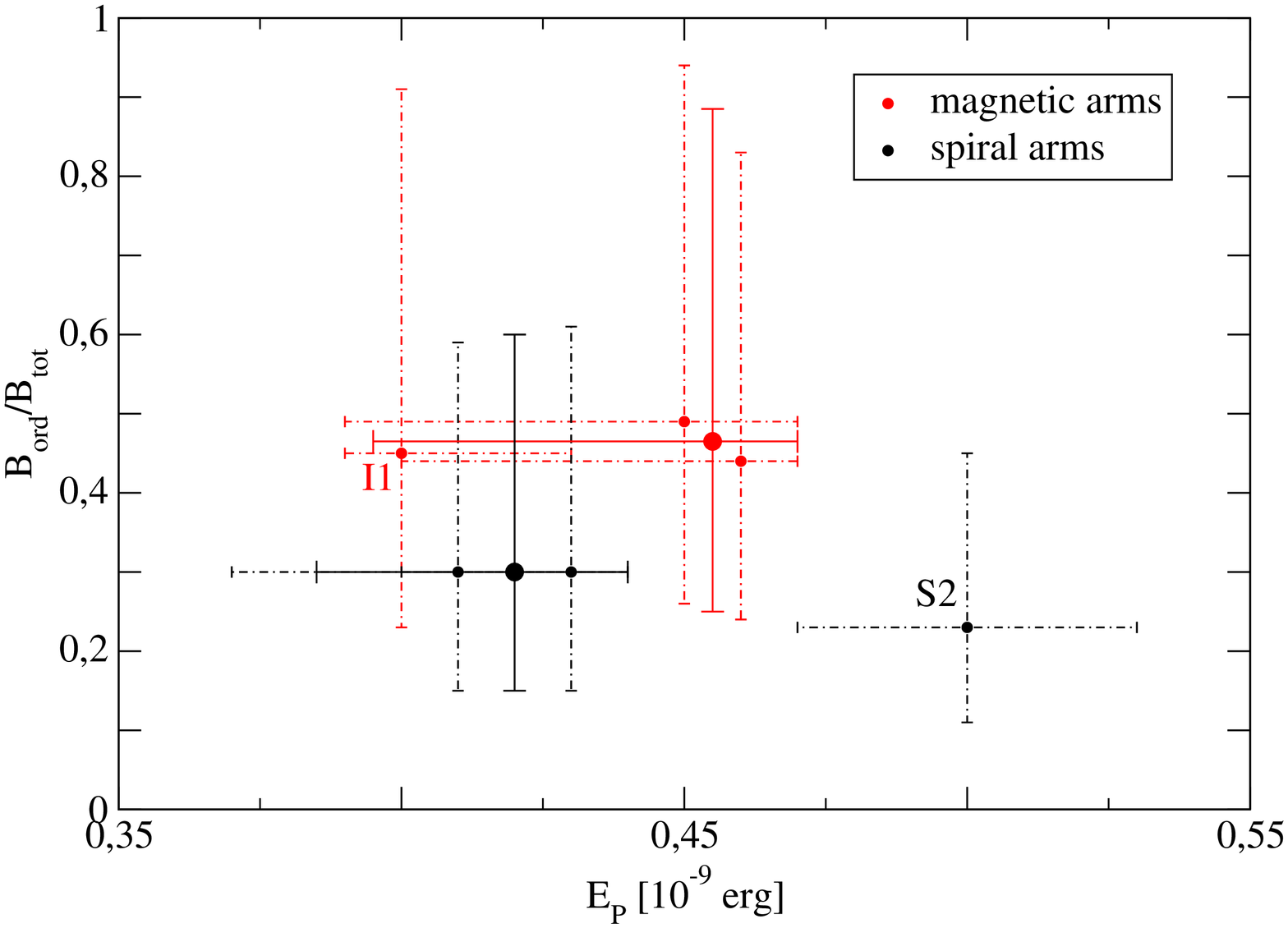}} 
	\caption{Relation between energy per particle (${\rm E}_{\rm P}$) and energy density of the magnetic field (${\rm\epsilon}_{\rm B}$) (left) and 
	order (${\rm B}_{\rm ord}/{\rm B}_{\rm tot}$) of the magnetic field (right) in the spiral (black) and the magnetic (red) arms of M\,83. Larger symbols 
	with solid line error bars refer to average values excluding I1 and S2 (see text for details). The smaller symbols with dot-dash error bars present individual regions.
        }
\label{epmag}
\end{figure*}

Because the reconnection acts more efficiently on the turbulent component of the magnetic fields, we observe more ordered magnetic fields. This is a result of an increase in the ratio of ordered-to-turbulent fields. This is indeed observed in the interarm regions of M\,83 in the form of the `magnetic arms' - large-scale ordered magnetic fields. 
For all regions of M\,83 we checked for the possible relation between the energies per particle of the hot gas and the order of the magnetic fields (right panel of Fig.~\ref{epmag}). 
As expected, the higher order of the magnetic fields in the interarm regions accompanied by the higher energies per particle would agree with the scenario presented above. We note that 
for the reasons presented above, the values obtained for regions S2 and I1 should be treated with care. The error bars presented in both plots come from conservative assumptions, that is, variation of 
the parameters used for the calculations of the magnetic field strengths by 50\%. Therefore, more realistic uncertanties should be lower. 

The parameters of the hot gas are averaged over the entire volume of the region, that is, in the disc and the above halo, and the temperature relations are different 
when each of the thermal components is considered separately. Nevertheless, because the reconnection heating could also occur in the galactic halo \citep{raymond92}, the discussion of 
the globally averaged values of the hot gas temperature seems to be justified. Especially, that the magnetic field reconnection in the halo would lead to the amplification 
of the line-of-sight component of the magnetic field (studied via Rotation Measures), that could help the disc hot gas to reach the halo more easily. 
A slight increase in temperature of the halo above the magnetic arm regions (Table~\ref{halotemps}) is supporting this hypothesis. 
Because the magnetic fields in the disc are significantly stronger than that in the halo, we expect that also the reconnection effects would be much stronger. Therefore, 
the regular component of the halo magnetic field, even when amplified by reconnection, 
should not significantly contribute to depolarisation of the disc magnetic field, which is observed in the plane of the sky.

\begin{table}[ht]
\caption{\label{onetemps}Average temperatures of the hot gas for the spectral regions in M\,83.}
\centering
\begin{tabular}{cl}
\hline\hline
Region      & kT$_{avg}$\tablefootmark{a} [keV] \\
\hline
\vspace{5pt}
S1          & 0.180$^{+0.008}_{-0.007}$ \\
\vspace{5pt}
S2          & 0.232$\pm$0.013	        \\
\vspace{5pt}
S3          & 0.175$^{+0.004}_{-0.001}$ \\
\vspace{5pt}
I1          & 0.166$^{+0.013}_{-0.005}$ \\
\vspace{5pt}
I2          & 0.184$^{+0.016}_{-0.027}$ \\
\vspace{5pt}
I3          & 0.203$^{+0.011}_{-0.031}$ \\
\hline
\end{tabular}
\tablefoot{
\tablefoottext{a}{Normalisation-weighted averages of kT$_1$, kT$_2$, and kT$_3$ from Table~\ref{m83xtabr}.}
}
\end{table}

\subsection{Comparison with NGC\,6946}
\label{comparison}

A similar analysis to that presented in the previous sections was already performed for another grand-design face-on spiral galaxy, NGC\,6946 \citep{wezgowiec16}.
The energies per particle calculated for the interarm regions of that galaxy were significantly higher than the corresponding values for the spiral arm regions. This trend is still visible in M\,83, though the values agree within errors. Surprisingly to this result, the magnetic arms 
are similarly prominent in both galaxies. As a consequence, almost identical values of the strengths of total and regular magnetic fields in NGC\,6946 and in M\,83 were derived.
For NGC\,6946, a slight increase in the temperature of the hot gas was observed in the interarm regions, compared to the spiral arms. This we also observe in M\,83, however the differences are still within the derived uncertainties. 

It is important to note here, that the star-forming regions in NGC\,6946 were closely related to the spiral arms and the interarm regions remained relatively free from any H$\alpha$ emission. This is different in M\,83, where the correspondence of the magnetic arms and the interarm regions is less clear (see above for the case of region I1). 
It seems that this property of a spiral galaxy is crucial to reliably test the hypothesis of a possible heating of the gas by the magnetic reconnection processes.

Another problem arises when a comparison sample is considered. It would be certainly important to compare our results with the observations of typical spiral galaxies that do not present 
any form of the magnetic arms, that is, in which the magnetic field structure is aligned with the spiral one. Surprisingly, all spiral galaxies that can be a subject of the study presented 
in this paper (thus having large angular exent and low inclination) show distinct magnetic arms \citep[see Table~6 in][]{beck19}. 
It is possible, that this is a common property of disc spiral galaxies. Nevertheless, 
sensitivities and resolutions provided by present radio telescopes do not allow to clearly detect the shift between the magnetic and the spiral structures in a galaxy of an angular extent of few arcminutes or smaller.  

\section{Summary and conclusions}
\label{cons}

In this paper we presented our analysis of the spectral properties of the hot gas in the disc and the halo of the spiral face-on galaxy M\,83, as well as the properties of its magnetic fields.
We compared the properties of the regions that are related to the star-forming spiral arms with that associated with the highly ordered magnetic fields in the interarm regions. Our results can be summarised as follows.

\begin{itemize}
\item[-] The very high sensitivity of the X-ray data for M\,83 allowed for the first time to observe two thermal components in the halo of a {\em face-on} galaxy. The derived associated temperatures 
correspond very well to earlier studies of edge-on spiral galaxies.
\item[-] The use of the Faraday depolarisation map of M\,83 allowed to estimate the extent of the gaseous halo, which was proven by the consistency of the derived parameters of the hot gas with that 
obtained for another spiral face-on galaxy NGC\,6946. These results are also confirmed by the star-forming properties of both galaxies.
\item[-] The spectral analysis of the hot gas in the spiral arms and the interarm regions of M\,83 yielded similar results to that for NGC\,6946, however for NGC\,6946 magnetic reconnection
as a proposed heating mechanism of the gas seems to work more efficiently.
\item[-] Our analysis suggests that the reconnection heating can indeed take place in and above the interarm regions of a spiral galaxy, but this effect is very difficult to observe and to constrain 
with the current sensitivity of the X-ray data. Especially important is the high contrast between the spiral arms and the interarm regions, when considering the H$\alpha$ star forming regions.
NGC\,6946 and M\,83 significantly differ in this aspect.
\end{itemize}
		
Observations of a larger number of similar spiral galaxies are needed to further verify the proposed hypothesis of the reconnection heating. Unfortunately, the required spatial and spectral 
sensitivity of the X-ray data, as well as the star-forming properties of a galaxy limit the possibility to perform studies presented in this paper to few objects only. 

\begin{acknowledgements}
The authors wish to thank the referee Robi Banerjee for suggestions that helped to improve the clarity of this paper. 
M.W., M.S., and M.U. are supported by the National Science Centre, Poland, with the grant project 2017/27/B/ST9/01050.
\end{acknowledgements}

\bibliographystyle{aa} 
\bibliography{myreferences} 

\end{document}